\documentclass[sigconf]{acmart}

\AtBeginDocument{%
  }

\acmConference[ASE'24]{The 39th IEEE/ACM International Conference on Automated Software Engineering}{Sacramento, California}{USA}

\acmISBN{978-1-4503-XXXX-X/18/06}

\usepackage{url}
\usepackage{amsmath,amssymb,amsfonts}
\usepackage{algorithm}
\usepackage{array}
\usepackage[noend]{algpseudocode}
\usepackage{graphicx}
\usepackage{wrapfig}
\usepackage{lipsum}
\usepackage{booktabs}
\usepackage{threeparttable}
\usepackage{subcaption}
\usepackage{textcomp}
\usepackage{enumerate}
\usepackage{paralist}
\usepackage{xcolor}
\usepackage{xspace}
\usepackage{multirow}
\usepackage{comment}
\usepackage{todonotes}
\usepackage{bbm}
\usepackage{nicefrac}
\usepackage{fancybox}
\usepackage{tcolorbox}
\hypersetup{
colorlinks = true, 
linkcolor=blue, 
citecolor=blue, 
urlcolor=blue 
}

\newcommand{\myparagraph}[1]{\smallskip\noindent\textbf{#1}\xspace}

\newcommand{\desInitialState}{\ensuremath{D_{\mathsf{I}}}\xspace}

\newcommand{\desIPS}{\ensuremath{D_{\mathsf{P}}}\xspace}

\newcommand{\testcase}{\ensuremath{\mathrm{S}}\xspace}

\newcommand{\initialState}[1]
{\ensuremath{\mathsf{I}_{#1}}\xspace}

\newcommand{\man}[1]
{\ensuremath{\mathsf{A}_{#1}}\xspace}

\newcommand{\ltcNo}[1]{\ensuremath{\mathsf{LS}_{#1}}\xspace}

\newcommand{\ctcNo}[1]{\ensuremath{\mathsf{CS}_{#1}}\xspace}

\newcommand{\pop}{\ensuremath{\mathcal{P}}\xspace}

\newcommand{\dangerSet}{\ensuremath{\mathcal{V}}\xspace}

\newcommand{\acr}{\ensuremath{f_{\mathsf{ACR}}}\xspace}

\newcommand{\mhd}{\ensuremath{f_{\mathsf{MHD}}}\xspace}

\newcommand{\dv}{\ensuremath{f_{\mathsf{DIV}}}\xspace}

\newcommand{\npcVehicleSet}{\ensuremath{V_{\mathsf{N}}}\xspace}

\newcommand{\npcVehicle}{\ensuremath{\mathsf{v}}\xspace}

\newcommand{\Legend}{\textsf{LeGEND}\xspace}
\newcommand{\llmone}{\ensuremath{\mathsf{LLM_1}}\xspace}

\newcommand{\llmtwo}{\ensuremath{\mathsf{LLM_2}}\xspace}
\newcommand{\legendminus}{\ensuremath{\Legend^{-}}\xspace}

\newcommand{\random}{\textit{Random}\xspace}

\newcommand{\avfuzzer}{\textit{AV-Fuzzer}\xspace}

\newcommand{\rom}[1]{\uppercase\expandafter{\romannumeral#1}}

\DeclareRobustCommand{\svdots}{
  \vbox{%
    \baselineskip=0.33333\normalbaselineskip
    \lineskiplimit=0pt
    \hbox{.}\hbox{.}\hbox{.}%
    \kern-0.2\baselineskip
  }%
}

\usepackage[switch]{lineno}

\begin{document}
\title[LeGEND: A Top-Down Approach to Scenario Generation of ADS Assisted by LLM]{LeGEND: A Top-Down Approach to Scenario Generation of Autonomous Driving Systems Assisted by Large Language Models}

\author{Shuncheng Tang}
\orcid{0000-0002-3019-2598}
\affiliation{%
  \institution{University of Science and Technology of China}
  \city{Hefei}
  \country{China}
}
\email{scttt@mail.ustc.edu.cn}

\author{Zhenya Zhang}
\orcid{0000-0002-3854-9846}
\affiliation{
\institution{Kyushu University}
\city{Fukuoka}
\country{Japan}
}
\email{zhang@ait.kyushu-u.ac.jp}

\author{Jixiang Zhou}
\orcid{0000-0002-4289-5151}
\affiliation{
\institution{University of Science and Technology of China}
\city{Hefei}
\country{China}
}
\email{zjxmail@mail.ustc.edu.cn}

\author{Lei Lei}
\orcid{0009-0000-4290-7549}
\affiliation{
\institution{University of Science and Technology of China}
\city{Hefei}
\country{China}
}
\email{lily168@mail.ustc.edu.cn}

\author{Yuan Zhou}
\orcid{0000-0002-1583-7570}
\affiliation{
\institution{Zhejiang Sci-Tech University}
\city{Hangzhou}
\country{China}
}
\email{yuanzhou@zstu.edu.cn}

\author{Yinxing Xue}
\orcid{0000-0002-2979-7151}
\authornote{Yinxing Xue is the corresponding author.}
\affiliation{%
  \institution{University of Science and Technology of China}
  \city{Hefei}
  \country{China}
}
\email{yxxue@ustc.edu.cn}

\begin{abstract}
  
\end{abstract}



\begin{abstract}
\emph{Autonomous driving systems (ADS)} are safety-critical and require comprehensive testing before their deployment on public roads. While existing testing approaches primarily aim at the criticality of scenarios, they often overlook the diversity of the generated scenarios that is also important to reflect system defects in different aspects. To bridge the gap, we propose \Legend, that features a top-down fashion of scenario generation: it starts with abstract functional scenarios, and then steps downwards to logical and concrete scenarios, such that scenario diversity can be controlled at the functional level.  
However, unlike logical scenarios that can be formally described, functional scenarios are often documented in natural languages (e.g., accident reports) and thus cannot be precisely parsed and processed by computers. To tackle that issue, \Legend leverages the recent advances of large language models (LLMs) to transform textual functional scenarios to formal logical scenarios. To mitigate the distraction of useless information in functional scenario description, we devise a two-phase transformation that features the use of an intermediate language; consequently, we adopt two LLMs in \Legend, one for extracting information from functional scenarios, the other for converting the extracted information to formal logical scenarios.
We experimentally evaluate \Legend on Apollo, an industry-grade ADS from Baidu. Evaluation results show that \Legend can effectively identify  critical scenarios, and compared to baseline approaches, \Legend exhibits evident superiority in diversity of generated scenarios. Moreover, we also demonstrate the advantages of our two-phase transformation framework, and the accuracy of the adopted LLMs.
\end{abstract}

\maketitle

\section{Introduction}
\label{sec:introduction}
\emph{Autonomous driving systems (ADS)} have been recognized as a revolutionary advancement in the automotive industry, offering a new solution to reducing mistakes by human drivers and mitigating traffic congestion. These systems bring unprecedented driving experience and significant efficiency, by employing intelligent sensors to perceive surrounding environments and sophisticated algorithms to make control decisions. However, despite these advantages, malfunctions of ADS are extremely dangerous, potentially leading to catastrophic consequences that pose fatal threats to human lives~\cite{tesla_death}. To ensure the safety of ADS, systematic testing is an indispensable stage before their deployment on public roads.

Given that real-world testing of ADS is prohibitively expensive, scenario-based testing in simulation environments has been widely adopted, thanks to its superior flexibility~\cite{tang2023survey}. The goal of this approach is to generate critical scenarios, e.g., in which ADS collide with other vehicles, so as to expose potential system defects and assist engineers to seek for effective remedies. The state-of-the-practice~\cite{li2020av, zhong2022neural, tian2022mosat, tang2023evoscenario, zhou2023collision, cheng2023behavexplor, huai2023doppelganger} often starts with a \emph{logical scenario} that specifies environments (e.g., road structure, weather) and traffic participants, but leaves a state space open (identified by a number of variables, e.g., initial states of vehicles), such that critical \emph{concrete scenarios} can be detected there by optimization-based search. 

While these approaches are effective to detect critical scenarios, they suffer from a severe issue about the \emph{diversity} of detected scenarios, i.e., the detected scenarios may exhibit similar behaviors of ADS that reflect similar system defects.  Even though some approaches~\cite{tian2022mosat, cheng2023behavexplor, zhou2023collision} mitigate this issue by taking diversity of scenarios as a search objective, the effect is still limited, since the detected scenarios remain constrained by logical scenarios.

\myparagraph{Motivations and Challenges} Compared to logical level, \emph{functional} level offers even more abstraction of scenarios that captures  their conceptual features only. For instance, in Fig.~\ref{fig:motivation_example}, while the logical scenario has strict restrictions on the space in which concrete scenarios can vary, functional scenarios just give a conceptual description about the featured events, involving the key actions and the interactions within the scenarios. To that end, starting scenario generation at the functional level is a plausible way to facilitate a great diversity of generated scenarios.

However, a challenging issue  arises in that direction: unlike logical scenarios that can be formally represented by \emph{domain-specific languages (DSL)}, functional scenarios are described conceptually at a more abstract level; consequently, they are often documented informally in natural languages. Given that the final objectives (i.e., concrete scenarios) of scenario-based testing are often formally characterized, we need to devise a systematic approach that can accommodate natural language inputs of functional scenarios.

\myparagraph{Contributions} We propose an approach \Legend (\underline{\textbf{L}}arge Language Model \underline{\textbf{E}}nabled \underline{\textbf{G}}eneration of Sc\underline{\textbf{EN}}ario for Testing of the Autonomous \underline{\textbf{D}}riving Systems) that exploits the recent advances of \emph{large language models (LLMs)} to handle natural language inputs, and can generate scenarios that are not only critical but also diverse. 
Specifically, \Legend adopts a top-down scheme that generates scenarios across three different abstraction layers of scenarios: 
it starts with abstract functional scenarios documented in natural languages and transforms them to logical scenarios that can be represented in domain specific languages; then, it searches in the open space of logical scenarios for critical concrete scenarios. Thanks to the top-down scheme, \Legend can effectively control, and thereby achieve a great diversity of generated scenarios.

To avoid unrealistic scenarios at the functional level, \Legend selects functional scenarios from real-world accident reports~\cite{nhtsa}. However, this raises another concern about the precision of LLMs, because there could be much distracting information in those accident reports. To that end, instead of using LLM directly, we
devise a two-phase transformation with the assistance of an intermediate representation, called \emph{interactive pattern sequence (IPS)},
that can record the featured events and their logical relations in reports. Consequently, \Legend involves two LLMs, i.e., \llmone that extracts featured events in accident reports into IPS, and \llmtwo that converts IPS to logical scenarios in formal DSL representations. 

We evaluate \Legend on Apollo~\cite{apollo}, which is an industry-grade ADS from Baidu. Experimental results show that, compared with two baseline approaches, \Legend can generate a more diverse set of critical scenarios. We also study the effectiveness of the intermediate representation (i.e., IPS), and the results show that it is indeed useful to improve the precision of transformation from natural language to formal scenarios. To understand the performance of the two LLMs adopted in \Legend, we also perform a user study that asks users to rate the performances of the LLMs in their respective tasks. Our study receives positive feedback about the performance of LLMs, which signifies their strengths in accomplishing these tasks and great potential in ADS testing.

In summary, this paper makes the following contributions:
\begin{compactitem}[$\bullet$]
    \item We propose \Legend, a top-down scenario generation approach that can achieve both criticality and diversity of scenarios;
    \item We devise a two-stage transformation, by using an intermediate language, from accident reports to logical scenarios; so, \Legend involves two LLMs, each in charge of one different stage;
    \item We implement \Legend and demonstrate its effectiveness on  Apollo, and we detect 11 types of critical concrete scenarios that reflect different aspects of system defects. All experiment results and the source code are available in~\cite{replication_package}.
\end{compactitem}

\section{Background}
\label{sec:background}
We first introduce scenarios and their abstraction hierarchy, and then we review  the scenario-based testing approach of ADS. 

\subsection{Abstraction Hierarchy of Scenarios}

Scenario is a commonly-used term in ADS testing. According to Ulbrich et al.~\cite{ulbrich2015defining}, it refers to a collection of actors (including ADS and other traffic participants, e.g., \emph{non-player character (NPC)} vehicles and pedestrians) attached with goals, actions and events, and their function environment (e.g., road structure, weather.).

As defined by Menzel et al.~\cite{menzel2018scenarios}, based on their levels of abstraction, scenarios can be categorized into \emph{functional scenarios}, \emph{logical scenarios}, and \emph{concrete scenarios}, as explained as follows:
\begin{compactitem}[$\bullet$]
    \item \emph{Functional scenarios} involve semantic descriptions about the participant entities and their interrelations; 
    \item \emph{Logical scenarios} provide  parameters and their respective ranges to characterize the state space of a functional scenario;
    \item  \emph{Concrete scenarios} are identified by assigning concrete values for each of the parameters in a logical scenario.
\end{compactitem}
 These three levels of abstraction offer a structured and systematic way that facilitates the safety assurance practice throughout the development of ADS~\cite{menzel2019functional}. More details can be found in~\S{}\ref{subsec:scenrio-based_ADS_testing}. 

\begin{figure}[!tb]
  \centering
  \includegraphics[width=0.95\columnwidth]{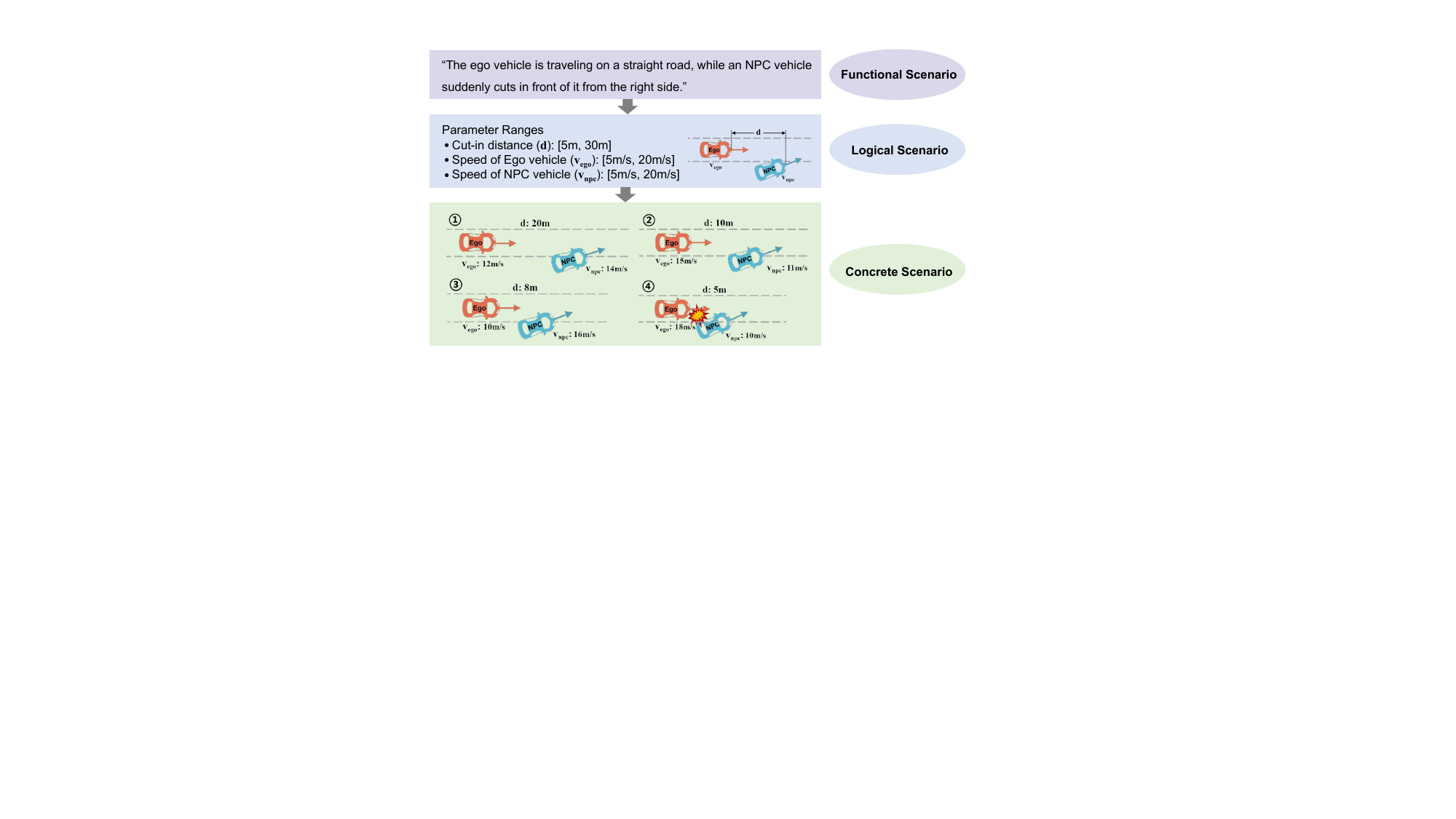}
  \caption{Scenarios in different abstraction levels}
  \label{fig:motivation_example}
\end{figure}
We present an example in Fig.~\ref{fig:motivation_example} to illustrate the scenarios in different abstraction levels.
As shown in Fig.~\ref{fig:motivation_example},  a functional scenario is a conceptual description about a scenario: \textit{``the ego vehicle is traveling on a straight road, while an NPC vehicle suddenly cuts in front of it from the right side''}. This functional scenario features an action \textit{``cut-in from right side''} from the NPC vehicle; however, it can still vary w.r.t. many factors, e.g., \emph{the cut-in distance} and \emph{the respective speed of the ego and NPC vehicles when cut-in happens}.

By selecting parameters in a functional scenario and fixing a range for each of them, we can obtain a logical scenario that constrains the variation space of scenarios (i.e., they can only vary w.r.t. the selected parameters and within the specified ranges). In Fig.~\ref{fig:motivation_example}, the logical scenario is obtained by selecting three parameters in the functional scenario and specifying a range for each of them.  

Then, by assigning different values for the parameters in a logical scenario, we can identify different concrete scenarios. A concrete scenario can be executed in a software simulator. Essentially, the aim of ADS testing is to find such concrete scenarios that can expose dangerous behaviors (e.g., collision with other vehicles) of ADS.

\subsection{Scenario-Based Testing of ADS}
\label{subsec:scenrio-based_ADS_testing}
Scenario-based testing has been a widely-adopted approach for safety assurance of ADS~\cite{tang2023survey}. To expose dangerous behaviors of ADS, it runs a variety of scenarios in software simulators and assesses their safety performances. In a plethora of literature~\cite{li2020av, zhong2022neural, tian2022mosat, tang2023evoscenario, zhou2023collision, cheng2023behavexplor, huai2023doppelganger}, scenario-based testing starts with a logical scenario that has a set  of searchable parameters $P$ concerning with the states of traffic participants, e.g., \emph{initial states of the ego vehicle} and \emph{actions of NPC vehicles}. Then, the search problem for critical scenarios can be cast to the following (multi-objective) optimization problem:
\begin{align}\label{eq:multiObjOpt}
    P^* = \arg\!\min_{P} \begin{cases}
        \mathcal{F}_1(P) \\
        \;\svdots \\
        \mathcal{F}_N(P)
    \end{cases} 
\end{align}
where $\mathcal{F}_i$ ($i\in\{1, \ldots, N\}$) is an objective function (a.k.a. \emph{fitness} function) that formalizes a specific search goal, including not only metrics for assessing the criticality of the scenarios, such as the minimal distance between vehicles and minimal \emph{time-to-collision}~\cite{tian2022mosat}, but also specific user preferences about scenarios, e.g., Calo et al.~\cite{calo2020generating} prefer the collisions that are avoidable.

The optimization problem aims to search for an optimal parameter  $P^*$ that minimizes the values of the objective functions. However, it may not be feasible to optimize all the objective functions at the same time, because these different search objectives may conflict with each other, in the sense that the decrease of one function may lead to the increase of another function. To that end, solvers for multi-objective optimization often aim to search for a \emph{Pareto-front} that consists of multiple solutions, each of which is optimal in terms of at least one objective function. There have been many established solvers, such as the widely-adopted one NSGA-II~\cite{deb2002fast}.

As explained in \S{}\ref{sec:introduction}, diversity of generated scenarios is an important property to pursue in ADS testing, in order to expose different types of system defects and avoid redundant exploration of search space. However, the existing approaches in Eq.~\ref{eq:multiObjOpt} often fail to generate scenarios with high diversity, because they do not explicitly take it into account in the design of testing approaches. Some techniques~\cite{tian2022mosat, cheng2023behavexplor, zhou2023collision} take a measure to mitigate this issue, by setting one of the objective functions to be a diversity metric; while it  helps to improve diversity, the effect is rather limited because the diversity of scenarios is still bounded by the high-level logical scenarios.

\begin{figure*}[!tb]
  \centering
  \includegraphics[width=0.8\textwidth]{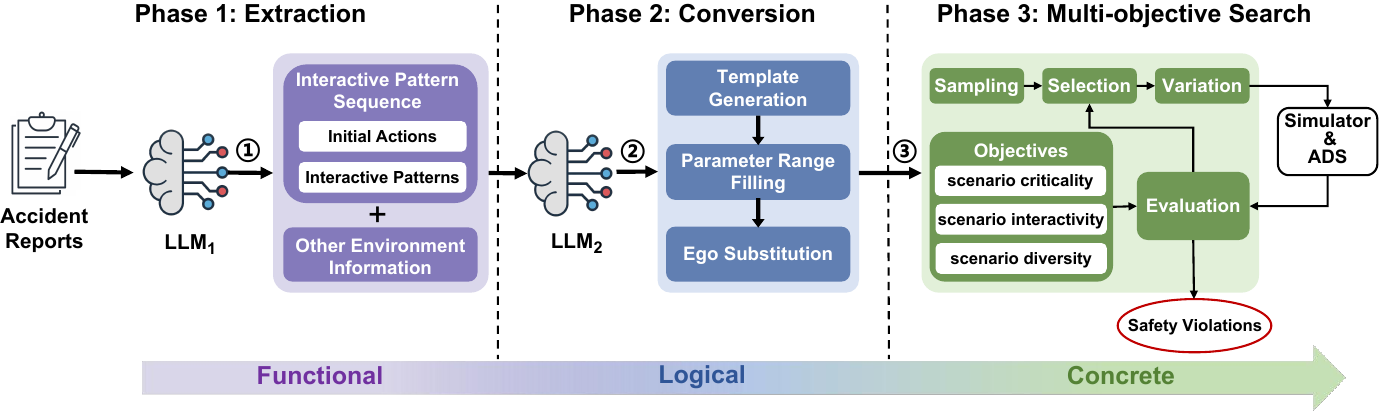}
  \caption{The workflow of \Legend}
  \label{fig:approach_overview}
\end{figure*}
\section{Approach Overview}
\label{sec:approach_overview}
In this paper, we propose an approach \Legend to scenario generation of ADS, which aims at not only criticality, but also diversity of generated scenarios. \Legend features a top-down style of scenario generation: unlike existing approaches that start scenario generation with the logical level, \Legend starts with the functional level  that is more abstract and thereby covers a much broader range of scenarios naturally, and then steps downwards to the logical level to search for critical concrete scenarios. In this way, \Legend can explore the scenario space at an abstract level and thereby hold a high-level control over the diversity of scenarios.

Unlike logical scenarios that can be expressed formally by DSL, functional scenarios are often documented informally by natural languages. As a consequence, it raises a challenging issue about the transformation from natural language descriptions to logical scenarios in formal DSL, where the latter are necessary for computers to automatically parse and search for concrete scenarios. To tackle this issue, we leverage the recent advances of \emph{large language models (LLMs)}, which have demonstrated unprecedented performances in cognition, logical reasoning, and natural language comprehension.

\myparagraph{Overview of \Legend}
The workflow of \Legend is depicted in Fig.~\ref{fig:approach_overview}, which consists of three phases. In this workflow, the first two phases are in charge of the transformation from natural language that documents functional scenarios to logical scenarios in formal DSL, and the last phase employs a search-based technique similar to existing studies (see~\S{}\ref{subsec:scenrio-based_ADS_testing}).  In the following, we elaborate on the design of the first two phases, as they are the main contributions.

To avoid unrealistic scenarios that may never happen in the real world, \Legend adopts functional scenarios documented in real accident reports. Specifically, we take these reports from~\cite{nhtsa}, which are official documents used to record the real accidents that happened in the United States during the years 2005-2007. The contents of these reports describe the factual information related to accidents such as car crashes, with detailed time, location, traffic participants, and the series of events happening before the crash. 

In the transformation of these reports to logical scenarios, a challenging issue arises about  the precision of  the transformation, since the reports are often lengthy and may contain considerable irrelevant information.
To that end, instead of using one LLM directly, we adopt an intermediate representation, called \emph{interactive pattern sequence (IPS)},  to record the featured events in the report, including the key actions and the interactions within the scenarios, and the logical relations between the events. In our empirical evaluation (in~\S{}\ref{subsec:rq2}), we show that IPS indeed play an important role in improving the precision of transformation, by a comparison with a straightforward transformation without using IPS. As a result, we employ two LLMs in the first two phases of \Legend, as follows: 
\begin{compactitem}[$\bullet$]
    \item In Phase 1, \llmone extracts the useful information in an accident report and records it into an IPS;
    \item In Phase 2, taking an IPS as input, \llmtwo translates it into a logical scenario represented in DSL.
\end{compactitem}

\begin{figure}[!tb]
     \centering
         \begin{subfigure}[b]{\linewidth}
         \centering
         \includegraphics[width=\linewidth]{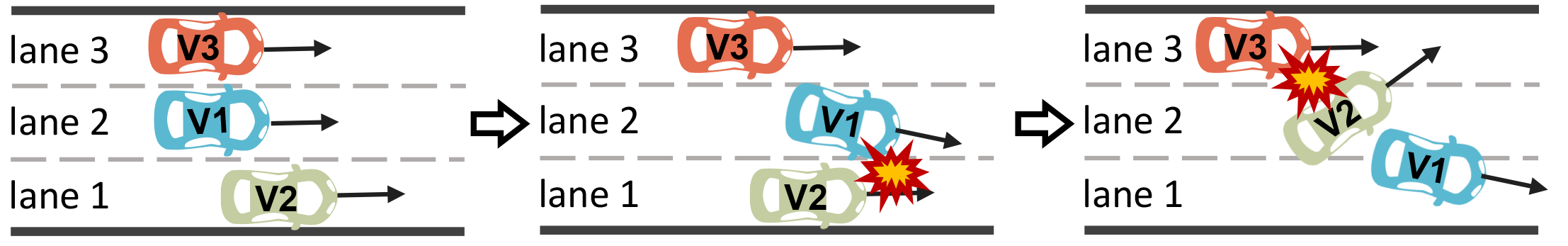}
         \caption{An illustration of a real accident~\cite{nhtsa_case}}
         \label{fig:reportIllustration}
         \vspace{1em}
     \end{subfigure}
     \begin{subfigure}[b]{\linewidth}
         \centering
         \includegraphics[width=\linewidth]{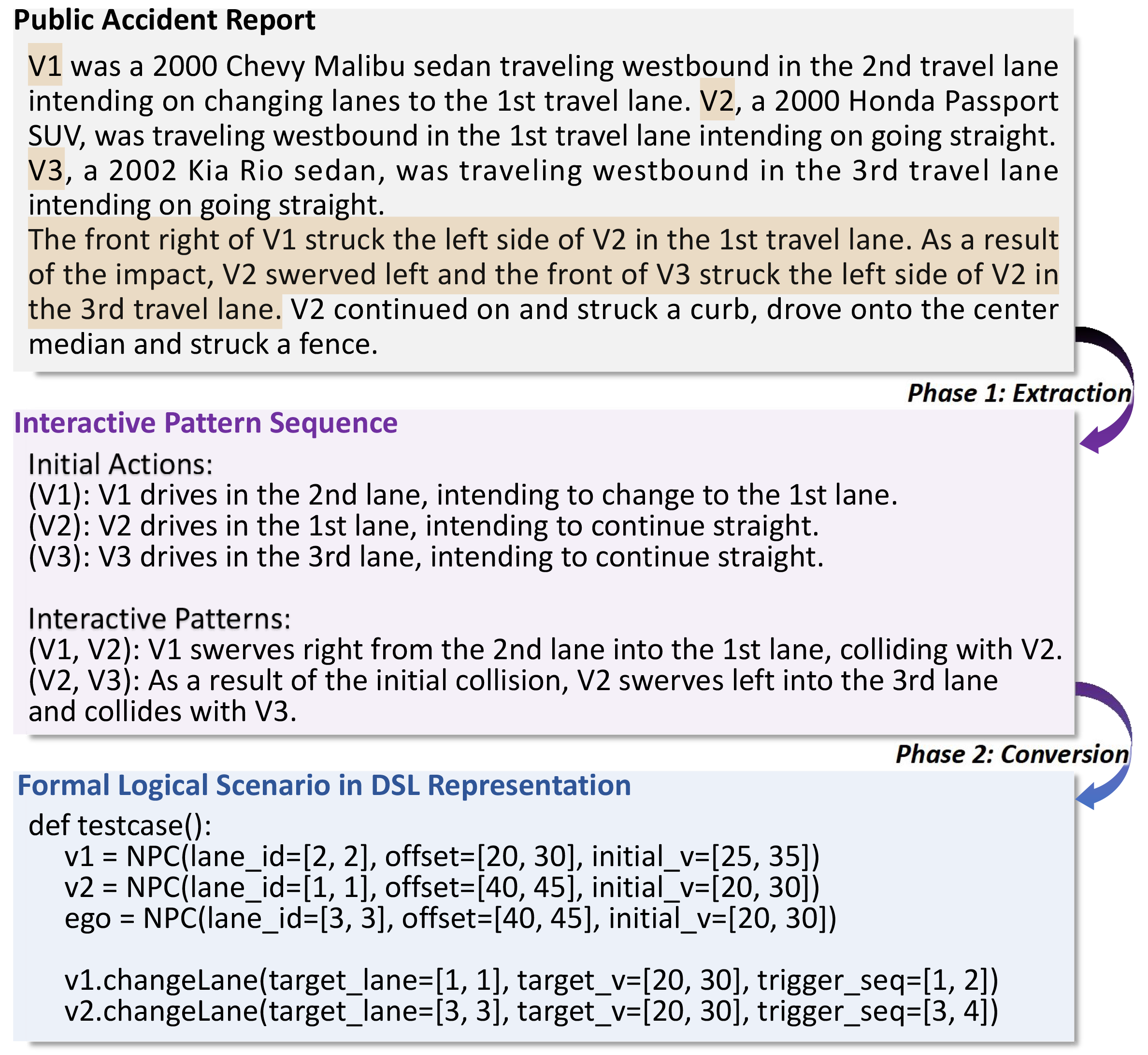}
         \caption{Transformation from the accident report to logical scenario}
         \label{fig:accidentReport}
     \end{subfigure}
        \caption{An example of transforming an accident report to a logical scenario in DSL}
        \label{fig:legend_example}
\end{figure}
\myparagraph{Example} We use an example, shown in Fig.~\ref{fig:legend_example}, to illustrate how \Legend transforms a real accident report to a logical scenario. The description of the accident is in the top plot of Fig.~\ref{fig:accidentReport}, with an illustration in Fig.~\ref{fig:reportIllustration}. The accident involves three vehicles, denoted as $V1$, $V2$ and $V3$, driving in three different lanes at the initial stage. Then, $V1$ wants to change to lane 1 by overtaking $V2$, but due to their close distance, $V1$ collides with $V2$; due to the impact of the collision, $V2$ swerves left and hits $V3$ in lane 3 subsequently.

In Phase 1, \llmone takes such an accident report and transforms it to an intermediate representation, shown as the middle plot of Fig.~\ref{fig:accidentReport}. In this  representation, we can see that it involves all necessary ingredients that characterize the accident, including the initial status of the vehicles, the interactions between vehicles that lead to collisions, and their  chronological relations embodied by the order of statements. This representation provides a concise format for recording the key information in accident reports, facilitating the subsequent transformation to formal representations. 

Having such an intermediate representation, in Phase 2, \llmtwo converts it to a logical scenario expressed by a formal DSL. Since the adopted DSL~\cite{tang2023evoscenario} also represents actions of vehicles in a sequential manner, \llmtwo can reliably accomplish the transformation. 

By these two phases, we obtain a logical scenario that faithfully reflects the ingredients in the functional scenario from real accident report, and so we can proceed with classic search techniques for detecting critical scenarios. In this loop, the diversity of scenarios can be controlled at the functional level, so we can devise our strategy to explore conceptually different scenarios on demand.

\section{Details of the Proposed Approach}
\label{sec:implementation}
In this section, we elaborate on the technical details of the three phases of \Legend, respectively in~\S{}\ref{subsec:FS_generation}, ~\S{}\ref{subsec:LS_conversion}  and~\S{}\ref{subsec:CS_identification}.

\subsection{Extraction from Accident Reports}
\label{subsec:FS_generation}
As introduced in~\S{}\ref{sec:approach_overview}, Phase 1 employs \llmone that takes an accident report as input and produces an interactive pattern sequence and other information such as road structures. We take a further look into the formats of the input and output of \llmone.

To avoid unrealistic scenarios that may never happen in the real world, we extract functional scenarios from public accident reports collected from the \emph{National Highway Traffic Safety Administration (NHTSA)} database~\cite{nhtsa}. These reports document real accidents that happened in the United States. The reports mainly consist of detailed descriptions about the environments, traffic participants, and a series of events (i.e., actions of different participants) preceding the accident. The accidents often take place on two types of road structures, i.e., straight and curved roads, and include multiple types of crashes, such as \emph{rear-end collisions} between multiple vehicles.

\llmone mainly outputs an intermediate representation called \emph{interactive pattern sequences}, which involve a tuple $\langle \desInitialState, \desIPS \rangle$, where
 $\desInitialState$ denotes the initial actions of vehicles, and $\desIPS$ describes the interactive patterns between vehicles before an accident. 
\begin{compactitem}[$\bullet$]
    \item $\desInitialState$ is a sequence of initial actions, with each action formatted as follows, where $V_{i}$ is a vehicle, followed by a natural language description about the states and the intended action of $V_{i}$.
    \begin{align*}
        \langle V_{i} \rangle: \{\text{initial states and actions of } V_{i}\}
    \end{align*}
    \item $\desIPS$ is a sequence of interactive patterns, with each pattern formatted as follows, where $V_{i}$ and $V_{j}$ are two vehicles, followed by a description about the actions of $V_{i}$ and $V_{j}$ w.r.t. their interaction:
    \begin{align*}
        \langle V_{i}, V_{j} \rangle: \{\text{actions of $V_{i}$ and $V_{j}$ related to interaction}\}
    \end{align*}
    Note that, the verbs used to describe actions are required to be selected from the set \textit{\{brake,\;decelerate,\;accelerate,\;swerve\;left/right\}}. Typically, an interactive pattern comes into being when one vehicle performs an active action and  another vehicle follows a responsive action. 
\end{compactitem}

\myparagraph{Prompt engineering} Interactive pattern sequences are generated via prompt engineering applied to the \llmone. Here, we provide an example to illustrate how \llmone is instructed for extracting an interactive pattern sequence from a given accident report. Specifically, we give the following extraction instruction to \llmone:
\begin{tcolorbox}[title = Prompt for \llmone, colback=white,coltitle=black, colframe=gray!20, arc=.3em, boxsep=-0.5mm]
\textit{\textbf{Task:} Please extract the road structure, and the interactive pattern sequence (IPS) from the given accident report. \\
\textbf{IPS Format:} \\
$V_i$: initial action of $V_i$. \\
$V_j$: initial action of $V_j$. \\
($V_i$, $V_j$): interactive actions between $V_i$ and $V_j$. \\
\textbf{Attentions:} $A1$, $A2$, ...} 
\end{tcolorbox}
The prompt applied to \llmone is structured into three parts: 
\begin{inparaenum}[1)]
    \item the first part specifies the task assigned to \llmone, namely, the extraction of  interactive pattern sequence and other information from the given accident report; 
    \item the second part provides a detailed template for the required  format of interactive pattern sequences, including the initial actions of the involved vehicles, followed by their interactive actions;
    \item  the third part outlines specific attentions that \llmone must adhere to, during the extraction, such as the requirement regarding the verbs they must select from the predefined set, as aforementioned. 
\end{inparaenum}

Moreover, the extracted interactive pattern sequences are subject to a legality check to ensure their correct formatting. The check is syntactic: for example, it checks whether there are exactly two vehicles participating in each interactive pattern. If the format is incorrect, \llmone is instructed to repeat the process. Once they can pass the legality check, these interactive pattern sequences are used as inputs for the subsequent phase of logical scenario conversion.

\subsection{Conversion to Logical Scenario }
\label{subsec:LS_conversion}
In Phase 2, the conversion from an interactive pattern sequence to a logical scenario involves three stages, i.e., template generation (\S{}\ref{subsubsec:template_generation}), parameter range filling (\S{}\ref{subsubsec:filling}), and ego substitution (\S{}\ref{subsubsec:ego_substitution}). 
Among these stages, the first two involve generation tasks, for which we utilize \llmtwo, while the final stage can be executed locally without LLM.
The outcome of this phase is a logical scenario in formal DSL, which can be automatically parsed and processed, as an input of Phase 3 for the search of concrete scenarios.

\subsubsection{Template Generation}
\label{subsubsec:template_generation} 
As introduced in~\S{}\ref{sec:background} and~\S{}\ref{sec:approach_overview}, a logical scenario can be formally represented in DSL. In this paper, we adopt a DSL in~\cite{tang2023evoscenario} that features a sequential representation of test cases (i.e., scenarios in our context), thus in line with the formats of interactive pattern sequences. We briefly introduce this DSL.
 
In this DSL, a test case \testcase is a sequence of initial states of vehicles and  actions of NPC vehicles, as follows:
\begin{align}\label{eq:testcase}
    \testcase = \{
    \underbrace{\initialState{1}, \ldots, \initialState{m}}_{\text{ initial states}}, \underbrace{\man{1}, \ldots, \man{n}}_{\text{ actions}}\}
\end{align}
where $\initialState{i}$ ($i\in\{1,\ldots, m\}$) is the initial state of the $i$-th vehicle, and $\man{j}$ ($j\in\{1,\ldots, n\}$) is an action of an NPC vehicle.
Specifically, the actions we consider include \emph{accelerating}, \emph{decelerating}, and \emph{lane changing}, each attached with different types of parameters.

From the perspective of implementation, both $\initialState{i}$ and $\man{j}$ are function calls, so they allow different types of parameters. The details of valid parameters for different actions are listed in Table~\ref{tab:testcase}. 

The task of \llmtwo in this stage is to convert an interactive pattern sequence to a \emph{template} of such a test case, namely, while the names of the functions in \testcase are fixed, the parameters are unspecified. The parameter ranges are later decided as presented in~\S{}\ref{subsubsec:filling}.

\myparagraph{Prompt engineering} Since the DSL we adopted is a formal language, it is necessary to ensure that \llmtwo can generate test cases with valid syntax. To that end, we employ \emph{one-shot learning} to instruct \llmtwo to accomplish the generation task. One-shot learning is a promising way~\cite{zhang2023multimodal} to train LLMs by prompt engineering: unlike cumbersome model training (e.g., \emph{fine-tuning} in LLM), one-shot learning can teach LLMs to learn desired output formats by feeding examples in prompts. In our case, since the format of interactive pattern sequences is close to the formalism of our adopted DSL, lightweight methods like one-shot learning could suffice.
 
Specifically, we give the following instruction to \llmtwo:
\begin{tcolorbox}[title = Prompt for \llmtwo, colback=white,coltitle=black, colframe=gray!20, arc=.3em, boxsep=-0.5mm]
\textit{\textbf{Task:} Please generate the test case template corresponding to the given functional scenario. \\
\textbf{Test Case Model \& Example:} $S$ \\
\textbf{Attentions:} $A1$, $A2$, ...} 
\end{tcolorbox}
Similar to the one in~\S{}\ref{subsec:FS_generation}, it also consists of a task instruction and a list of attentions; the difference mainly involves the example of the formalism of DSL, in the part of \textit{\textbf{Test Case Model \& Example}}. It provides the description of a test case, and a template in the form of Eq.~\ref{eq:testcase}, which can then be learned by \llmtwo. 

\subsubsection{Parameter Range Filling}
\label{subsubsec:filling}
In this stage, we fill the parameter ranges in the generated template. While this can be done by LLMs with information about road specifications and default parameters,
 by our observation (outlined in \S{}\ref{subsec:rq3}), the parameter ranges filled by LLMs may not be entirely rational. For instance, consider two vehicles initially positioned in the same lane; the LLM might generate parameter ranges like $[0, 5]$ and $[5, 10]$ for their positions. Since each vehicle has its own length, these ranges could result in a collision at the beginning of the simulation. Our approach of determining the parameter ranges is rule-based, and the idea is essentially that, if the parameter ranges generated by LLM are valid and more precise than default ones, we adopt them; otherwise, we stay conservative to adopt the default ones. These default ranges for parameters are collected from existing literature and map documentation and are often conservative. They are accessible via our repository (see~\S{}\ref{sec:experiment_design}).

\begin{table}[!tb]
  \small
  \centering
  \caption{Test Case Model}
  \label{tab:testcase} 
    \resizebox{\linewidth}{!}{
    \begin{tabular}{p{0.2\linewidth}  p{0.2\linewidth} p{0.6\linewidth}}
    \toprule
    \textbf{Type} & \textbf{Name} & \textbf{Parameter} 
    \\
    \midrule
     Constructor & NPC  
     & Lane id, lane offset, initial speed \\
     
     Method & Accelerate   
     & Target speed, trigger sequence \\
    
     Method & Decelerate   
     & Target speed, trigger sequence \\
    
    Method & Lane changing
    & Target lane, target speed, trigger sequence \\ 
    \bottomrule
    \end{tabular}}
\end{table}

\subsubsection{Ego Substitution}
\label{subsubsec:ego_substitution}
To apply the scenarios in ADS testing, we need to assign one vehicle involved as the ego ADS vehicle.
Given that an accident typically involves multiple vehicles, the ego vehicle can be  any one involved in the accident. 
To assess ADS in more challenging situations, we retain those vehicles that perform active actions (that finally lead to accidents) as NPCs, and substitute the one that is relatively passive in the accident as the ego vehicle.
To identify this vehicle, we analyze the interactive pattern sequences by counting the frequency of active actions performed by all involved vehicles. The vehicle with the lowest frequency of active actions is then selected as the ego vehicle. This selected NPC vehicle is then designated as the ego vehicle, and its actions are removed from the test case template to hand over control to  ADS algorithms.

\subsection{Search-Based Concrete Scenario Generation}
\label{subsec:CS_identification}
With logical scenarios generated from \S{}\ref{subsec:LS_conversion}, we leverage a multi-objective genetic algorithm, as presented in Alg.~\ref{alg:search}, to search for critical concrete scenarios. Overall, the algorithm consists of three main steps: \emph{initial sampling} (Line~\ref{line:1stGen}-Line~\ref{line:addNewTest}), \emph{pareto-optimal selection} (Line~\ref{line:ParetoOpt}), and  \emph{genetic variation} (Line~\ref{line:2ndGen}-Line~\ref{line:addMutatedCase}). Next, we introduce these steps in detail.

\begin{algorithm}[!tb]
    \caption{Multi-objective search algorithm}
    \footnotesize
    \label{alg:search}
    \renewcommand{\algorithmicrequire}{\textbf{Input:}}
    \renewcommand{\algorithmicensure}{\textbf{Output:}}
    \begin{algorithmic}[1]
    \Require Population size $p_{\mathit{max}}$, logical scenario $\ltcNo{}$ and size $s_{\mathit{max}}$, maximum number  of generations $g_{\mathit{max}}$
    \Ensure  Critical test case set $\dangerSet$   
    \State $\dangerSet \gets \emptyset$ \Comment{the set of critical scenarios}
    \State $\pop \gets \emptyset$ \Comment{the population during search}
    \For{$g \in \{1,\ldots, g_{\mathit{max}}\}$}
    \State $\pop' \gets \emptyset$
        \If{$g = 1$} \label{line:1stGen}\Comment{the 1st generation}
            \While{$\lvert \pop' \rvert < p_{\mathit{max}}$}
             \State $\ctcNo{} \gets \emptyset$
             \For{$i \in \{1,\ldots, s_{\mathit{max}}\}$}
             \State $\ctcNo{}(i) \gets \textsc{Fill}(\ltcNo{}(i))$ \label{line:fill}
             \State $\ctcNo{} \gets \ctcNo{} \cup \{\ctcNo{}(i)\}$ \label{line:testGen}
             \EndFor
             \State $\pop' \gets \pop' \cup \{\ctcNo{}\} $ \label{line:addNewTest}
            \EndWhile
        \Else \label{line:2ndGen}\Comment{from the 2$nd$ generation on}
        \While{$\lvert \pop' \rvert < p_{\mathit{max}}$} \label{line:checkVarCap}
            \State tournament selection of $\ctcNo{1}, \ctcNo{2}\in \pop$ \label{line:tournament}\Comment{selection}
            \State randomly select index $j$ \label{line:crossover}\Comment{crossover}
            \State$\ctcNo{1}', \ctcNo{2}'\gets$ swap the $j$th npc in $\ctcNo{1}, \ctcNo{2}$ \label{line:swap}
            \For{$i \in \{1,\ldots, s_{\mathit{max}}\}$} \label{line:mutation}\Comment{mutation}
              \State mutate parameters of $\ctcNo{1}'(i)$ and $\ctcNo{2}'(i)$
            \EndFor
            \State $\pop' \gets \pop' \cup \left\{ \ctcNo{1}', \ctcNo{2}' \right\}$ \label{line:addMutatedCase}\Comment{add $\ctcNo{1}', \ctcNo{2}'$ to $\pop'$}
            \Statex 
        \EndWhile
        \EndIf
        \State $\pop \gets \arg_{\ctcNo{}\in \pop\cup \pop'}
            \begin{cases}
            \min \mhd \\
            \max \acr \\
            \max \dv
            \end{cases}
        $ \label{line:ParetoOpt}\Comment{Pareto-optima} 
        \For{$\ctcNo{}\in \pop$} \label{line:criticalStart}
            \If{$\mhd(\ctcNo{}) < l $}  \Comment{critical scenario found}
            \State $\dangerSet\gets \dangerSet \cup \{\ctcNo{}\}$ \label{line:criticalReturn}\Comment{return critical ones}
            \EndIf
        \EndFor
    \EndFor
    \end{algorithmic}
\end{algorithm}

\subsubsection{Initial Sampling}
First, \Legend constructs an initial population set consisting of random samplings of concrete scenarios. By filling each statement in a logical scenario with a randomly generated parameter value sampled from the corresponding range (Line~\ref{line:fill}), it constructs a concrete scenario (Line~\ref{line:testGen}).
The generated concrete scenarios are then added into a temporary set (Line~\ref{line:addNewTest}).

\subsubsection{Pareto-optimal Selection}
\label{subsubsec:selection}
To guide the search, we adopt three objective functions, including \emph{scenario criticality}, \emph{scenario interactivity}, and \emph{scenario diversity}, adapted from literature~\cite{tian2022mosat, tang2023evoscenario}. Due to these different objective functions, we need to search for and select the Pareto-optimal solutions.
Below, we elaborate on the definitions of these objective functions, under the assumption that the simulation is executed over a time interval $[0, T]$. 

\myparagraph{Scenario Criticality}
Given that the primary objective is to identify critical scenarios that can expose safety violations, the first objective function is the \emph{Minimum Headway Distance} (\emph{MHD})~\cite{abuelenin2015effect}. This metric measures the minimum headway distance of the ego vehicle to other NPCs during the simulation. Given a set \npcVehicleSet of NPC vehicles, the \mhd is calculated as follows:
\begin{align*}
    \mhd \;:=\; \min_{t\in[0, T]}\min_{\npcVehicle\in \npcVehicleSet} \mathsf{Dis}(\mathsf{Head}_{\mathsf{ego}}, \mathsf{Head}_{\npcVehicle})
\end{align*}
where $\mathsf{Head}_{\mathsf{ego}}$ and $\mathsf{Head}_{\npcVehicle}$ represent the headway positions of the ego vehicle and an NPC vehicle $\npcVehicle$ respectively, and $\mathsf{Dis}$ calculates the Euclidean distance between the two positions. A smaller \mhd corresponds to a higher fitness score; when the \mhd falls below a threshold value $l$, such as the length of a vehicle, it indicates a collision between the ego vehicle and NPC vehicles, thereby we can classify it as a critical scenario (Line~\ref{line:criticalStart}-Line~\ref{line:criticalReturn}).

\myparagraph{Scenario Interactivity}
With this objective, we aim to produce critical scenarios by enhancing the interactivity between the ego vehicle and NPC vehicles. We adopt the metric \emph{Acceleration Change Rate} (\emph{ACR})~\cite{tian2022mosat}, which quantifies the interactivity between vehicles by measuring the rate of change in the ego vehicle's acceleration in response to the behaviors of NPC vehicles. \acr is as follows:
\begin{align*}
    \acr \;:=\;& \frac{1}{T} \left(\textstyle\sum_{t_1,t_2 \in \left[0, T\right]} \mathbbm{1} \big( \big\lvert a(t_1) - a(t_2) \big\rvert \geq \eta \big)\right) \\
     \text{s.t.} &\quad \dot{a}(t_1)=0, \quad \dot{a}(t_2)=0 
\end{align*}
where $a$ represents the acceleration of the ego vehicle, $\dot{a}$ denotes its first-order derivative, and $\mathbbm{1}$ is the indicator function with $\eta$ as a threshold. The indicator function returns 1 if the specified condition is met and 0 otherwise. The more frequent interactions are, the greater the value of \acr is. 

\myparagraph{Scenario Diversity}
This objective is to explore a wide range of parameter configurations under the same logical scenarios, thereby increasing the variety of generated scenarios. To that end, we adopt \dv, which measures the average Euclidean distances over generated scenarios, by treating each scenario as a point (the dimension of which is the same as the number of parameters in the logical scenario). A larger \dv value indicates a broader exploration in the parameter configuration space.

We then proceed to identify Pareto-optimal test cases based on their values across the three objectives. Specifically, all of the test cases that exhibit superior performance in at least one objective are retained within the population. These selected test cases will undergo a genetic variation phase, as detailed in \S{}\ref{subsubsec:variation}, to generate new candidate test cases for the subsequent generations. 

\subsubsection{Genetic Variation}
\label{subsubsec:variation}
From the second generation on (Line~\ref{line:2ndGen}), our algorithm, as illustrated in Alg.~\ref{alg:search}, constructs a candidate set of test cases based on those selected Pareto-optima in \S{}\ref{subsubsec:selection}. This process involves two types of genetic variation operations: \emph{crossover} and \emph{mutation}. These operations are designed to handle the sequential test case model, as described subsequently, and ensure effective exploitation by the algorithm.

\myparagraph{Crossover} 
In each iteration, this operation acquires two test cases $\ctcNo{1}$ and $\ctcNo{2}$ through tournament selection, based on the evaluation of the objective function (Line~\ref{line:tournament}). Subsequently, a vehicle index 
$j$ is randomly selected, and the $j$-th NPC vehicle, including its initial positions and driving maneuvers, is tentatively swapped between the two concrete test cases. This process will yield two new test cases $\ctcNo{1}'$ and $\ctcNo{2}'$ (Line~\ref{line:swap}). 

\myparagraph{Mutation} For each statement in $\ctcNo{1}'$ and $\ctcNo{2}'$, this process will apply a polynomial mutation~\cite{zhong2022neural} to each discrete and continuous variable. Note that the bounds for parameters have been fixed as presented in the logical scenarios.

\section{Experiment Design}
\label{sec:experiment_design}
 We implement \Legend as a Python library based on the simulation environment of Baidu Apollo~\cite{apollo}. All the code and experimental data are available in a public repository\footnote{\url{https://github.com/MayDGT/LeGEND}}.

\myparagraph{Research Questions}
To evaluate the performance of \Legend, we investigate the following research questions:

\begin{compactitem}[$\bullet$]
    \item \textbf{RQ1:} {\it Can \Legend improve diversity of scenarios, compared to baseline approaches?} In this RQ, we investigate the diversity of critical scenarios detected by \Legend. Moreover, we compare \Legend with baseline approaches under different metrics.

    \item \textbf{RQ2:} {\it Can interactive pattern sequences help to improve transformation precision?} In this RQ, we study the usefulness of  interactive pattern sequences for transformation precision.
    
    \item \textbf{RQ3:} {\it How accurate are \llmone and \llmtwo for the transformation tasks in \Legend?} In this RQ, we study the accuracy of \llmone for extraction in Phase 1 and \llmtwo for conversion in Phase 2. 
\end{compactitem}

\myparagraph{Baselines and metrics}
To address RQ1, we perform a comparative analysis with two baseline approaches: \random and \avfuzzer~\cite{li2020av}. The \random approach generates NPC vehicles and their driving actions randomly. \avfuzzer initially generates scenarios randomly, and then employs a genetic algorithm to search for more critical scenarios. However, \avfuzzer has limited ability to vary the number of NPC vehicles and select different road structures during testing. Therefore, for comparison purposes, we evaluate \avfuzzer in a fixed road environment with a constant number of NPC vehicles, such as a straight road with two NPC vehicles. 

We run \Legend, \random, and \avfuzzer for the same number of simulations and record the critical scenarios that contain collisions between ego and NPC vehicles. These critical scenarios are classified into different types based on their logical patterns, i.e., the actions of the vehicles prior to each collision (see~\S{}\ref{subsec:rq1} for classification results).
We then measure their performance in terms of the following metrics:
\begin{compactitem}[$\bullet$]
    \item {\it The number of distinct types of critical scenarios, denoted as} $\textsf{\#Types}$. This metric quantifies the total number of unique types of critical scenarios identified by different approaches.

    \item {\it The exposure rate of distinct types over the identified critical scenarios, denoted as} $\textsf{\#TypeExposRate}$. This metric measures the proportion of critical scenarios that reveal a distinct type of collision. It is calculated as the ratio of the number of distinct types to the total number of critical scenarios identified.
    
    \item {\it The number of simulations required to detect the first type of critical scenarios, represented by} $\textsf{\#SimForFirstType}$. This metric measures the number of simulations needed to identify the initial type of critical scenario.
    
    \item {\it The number of simulations required to detect all types of critical scenarios, denoted as} $\textsf{\#SimForAllTypes}$. This metric quantifies the number of simulations needed to uncover all distinct types of critical scenarios.
    
    \item {\it The time cost for one scenario, denoted as}  $\textsf{\#TimeForOneScenario}$. This metric evaluates the efficiency of different approaches by calculating the time duration for a single scenario, including its generation, execution, and analysis.
    
\end{compactitem}

To address RQ2, we conduct an ablation study to assess the usefulness of the extracted functional scenarios, especially the interactive pattern sequences, for the critical scenario generation process. For this purpose, we implement a variant of \Legend, called \legendminus. In \legendminus, only one  LLM is used to generate logical scenarios directly from accident reports, without extracting the functional scenarios. We run \legendminus and \Legend based on the same set of accident reports, and use the same number of simulations as the budget for searching concrete scenarios.
We then compare the performances of them using the metrics in RQ1. 

To address RQ3, we investigate the accuracy of \llmone for the extraction in Phase 1 and \llmtwo for the conversion in Phase 2. We answer RQ3 empirically by a user study, since it is difficult to assess the accuracy of the transformations. Specifically, we perform an online survey with 10 graduate students recruited from our department of computer science. Among the participants, six people have more than two years' research experience in the field of autonomous driving. In the study, since it takes one user about 5 minutes to finish the questions regarding one accident report, we randomly selected 5 accident reports from our database as test seeds. We then ask each participant to read each accident report and express their opinions on the statements regarding the extracted functional scenarios and converted logical scenarios. Participants are asked to rate their agreement on a 5-point Likert scale~\cite{likert1932technique}, ranging from ``Strongly Disagree'' to ``Strongly Agree''. In addition, we give participants an opportunity to provide supplementary feedback, such as brief textual comments, particularly for the instances on which their opinions are not strong.

\myparagraph{Configurations}
\Legend employs GPT4~\cite{achiam2023gpt} for both the extraction and conversion tasks, since this model is the state-of-the-art for a wide range of downstream tasks. We use the \emph{gpt-4-0613} checkpoint with $max\_token$ of 1000 provided via the OpenAI API~\cite{gpt4endpoint}, and the temperatures are all set as 0.8. To ensure realism of scenarios, we select 20 accident reports from the NHTSA database~\cite{nhtsa} as our initial seeds that cover different roads and numbers of vehicles. For the search algorithm, the crossover rate is set as $0.4$, and the mutation rate is set as $0.5$. Given that different test case templates vary in length (i.e., the number of statements in each test case), we set the population size as that length, following~\cite{lande1987effective}. The maximum number of iterations for generation is set to 10. Each driving action is executed for a duration of 5 seconds following the trigger sequence.
For RQ1 and RQ2, we run \Legend and baseline approaches for the same number of $1400$ simulations, and each execution lasts for more than 12 hours. For mitigating randomness, each experiment is repeated for 5 times and we take the average results.

Our experiments are conducted on a desktop with Ubuntu 20.04, 32GB of memory, an Intel Core i7-13700K CPU, and an NVIDIA RTX 3090 GPU. The simulation environment for the evaluation consists of Baidu Apollo 7.0~\cite{apollo} (the target ADS) and LGSVL 2021.3~\cite{rong2020lgsvl} (the software simulator). Apollo is an open-source and industry-grade modular ADS that supports a wide range of functionalities. We chose the San Francisco HD map provided by LGSVL, as this map includes various road structures that conform with the requirements about roads derived from the accident reports.

\section{Evaluation}
\subsection{RQ1: Can \Legend improve diversity of scenarios, compared to baseline approaches?}
\label{subsec:rq1}
\subsubsection{Qualitative Analysis}
\label{subsubsec:case_example}
During the execution, \Legend identifies 96 critical scenarios that reveal safety violations. These violations are further categorized into 11 distinct types based on the action patterns that lead to each collision. These types of critical scenarios encompass a diverse range of road structures, numbers of vehicles, and driving actions.
The distribution of critical scenarios across each type is shown in Fig.~\ref{subfig:violation_type}.
A comprehensive illustration of all types of critical scenarios is available on our website~\cite{replication_package}.
Due to space limitation, we select one type of critical scenario as an example to demonstrate that \Legend can identify critical scenarios derived from real-world interactive patterns. The evolution of this scenario is explained below.

\begin{figure}[!tb]
  \centering  
\includegraphics[width=\linewidth]{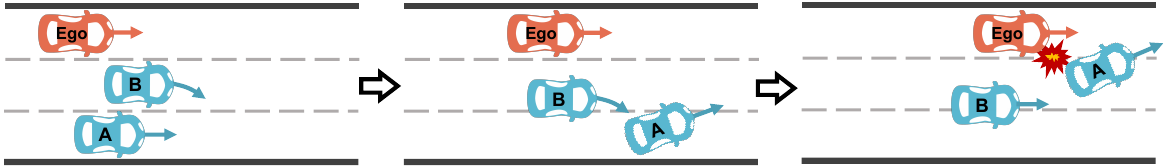}
  \caption{Example of a critical scenario derived from the accident report as illustrated in Fig.~\ref{fig:legend_example}}
  \label{fig:violation_example}
\end{figure}

As shown in Fig.~\ref{fig:violation_example}, the ego vehicle
is traveling on a straight road with two NPC vehicles, denoted as $A$ and $B$, driving in the adjacent two lanes. Suddenly, vehicle $B$ attempts to change lanes into the lane that vehicle $A$ drives and initiates a deceleration action. The ego vehicle detects the lane change of vehicle $B$ and responds by slightly braking before continuing to drive straight.
During vehicle $B$'s lane change, vehicle $A$ decides to move into the lane that the ego drives. However, the ego vehicle continues driving forward and fails to respond timely to vehicle $A$'s lane change, leading to a collision. 

Note that the logical pattern of this critical scenario is derived from a real-world accident report as discussed in \S{}\ref{sec:approach_overview}. The description of the report is shown in Fig.~\ref{fig:legend_example}. During the testing process, \Legend substitutes the ``V3'' vehicle with the ego vehicle controlled by the ADS to create a challenging driving environment.

\subsubsection{Quantitative Analysis}
\begin{figure}[!tb]
\begin{subfigure}{0.49\linewidth}
\centering
    \includegraphics[scale=0.3]{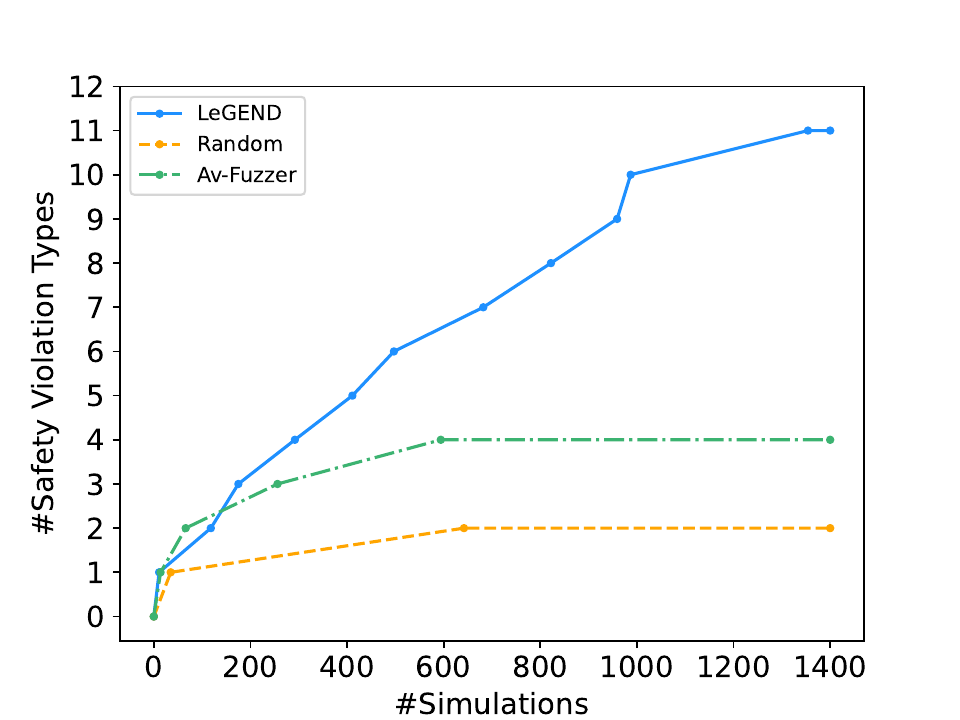}
    \caption{Safety violation types discovered over simulations}
    \label{subfig:comparison_types}
\end{subfigure}
\hfill
\begin{subfigure}{0.49\linewidth}
\centering
    \includegraphics[scale=0.3]{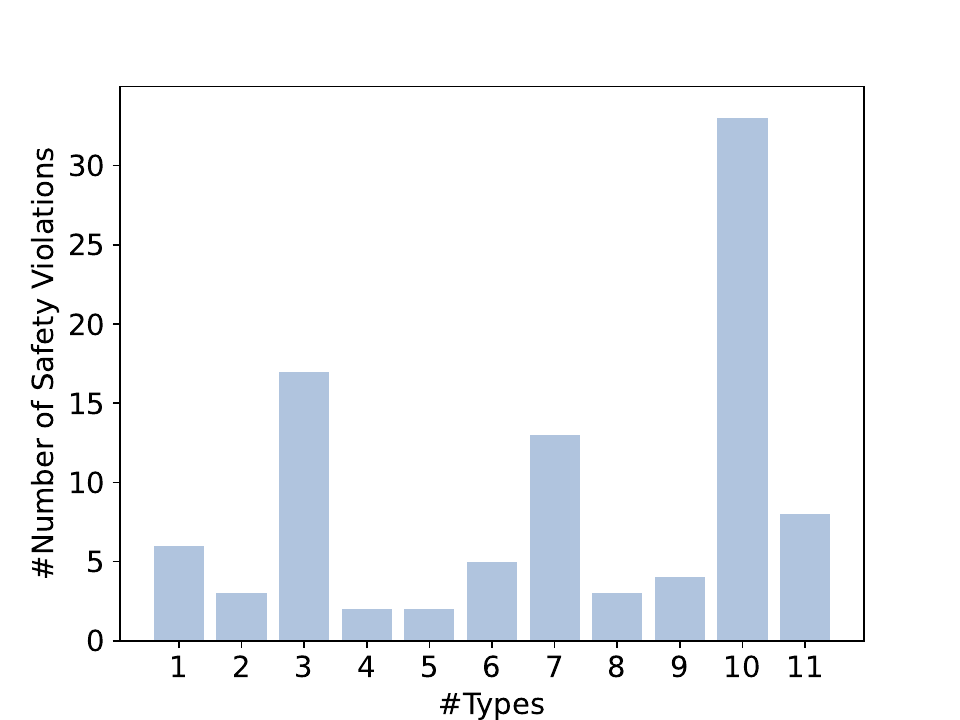}
    \caption{Number of safety violations of each type}
    \label{subfig:violation_type}
\end{subfigure}
\caption{The trend of violation types over simulations and the distribution of safety violations over each type}
\label{fig:type_violation}
\end{figure}

To further demonstrate the performance of \Legend, we compare it with the two baseline techniques based on the five metrics as outlined in \S{}\ref{sec:experiment_design}.
Fig.~\ref{subfig:comparison_types} illustrates the growth in the number of critical scenario types discovered by the three approaches over simulations. We can observe that \Legend outperforms the other methods since it manages to detect the highest number of distinct types of critical scenarios. Moreover, \Legend continuously exposes new types of critical scenarios throughout the simulations. In comparison, \avfuzzer exhibits the fastest convergence but detects only five distinct types of critical scenarios. \random exhibits a similar convergence speed to \avfuzzer but identifies the fewest types of critical scenarios. Notably, all types of critical scenarios detected by both \avfuzzer and \random are exposed within the first 700 simulations.

\textbf{\Legend vs. \random} Table~\ref{tab:comparison_with_baselines} presents a comparative analysis between \Legend and \random. Over the period of 1400 simulations, \random generates 64 critical scenarios that reveal potential system faults, with the first critical scenario identified during the 34-$th$ simulation. It can identify 2 types of critical scenarios during the scenario generation process. All types of the critical scenarios are observed within the first 642 simulations.

Compared to \random, \Legend generates a larger number of critical scenarios and uncovers a greater diversity of them. The type-exposure frequency $\textsf{\#TypeExposRate}$ of \Legend is higher, which indicates its ability to generate diverse critical scenarios effectively.
Additionally, the two types of critical scenarios identified by \random are also detected by \Legend in the 118-$th$ and 497-$th$ simulations, respectively. These comparative results demonstrate that \Legend is more effective and efficient in generating critical scenarios with greater diversity in terms of types of scenarios.

\textbf{\Legend vs. \avfuzzer} The comparison between \Legend and \avfuzzer is also detailed in Table~\ref{tab:comparison_with_baselines}. During the 12-hour execution, \avfuzzer generates 297 critical scenarios with the first collision occurring in the 14-$th$ simulation. The high number of identified critical scenarios by \avfuzzer can be attributed to the local search mechanism employed by this technique. However, despite the large number of critical scenarios, \avfuzzer is limited to exposing only 4 distinct types, resulting in a type exposure frequency of $1.3\%$.
The four types of critical scenarios are all exposed by \Legend in the 118-th, 411-$st$, 497-$th$ and 822-$nd$ simulation.

Compared to \avfuzzer, all four types of critical scenarios identified by \avfuzzer are also detected by \Legend. Moreover, \Legend exhibits a greater diversity of distinct critical scenarios during the critical scenario generation process. The type exposure rate for \Legend is $11.5\%$, significantly higher than the $1.6\%$ observed with \avfuzzer. Furthermore, \Legend detects the first critical scenario more quickly and has a lower time cost per scenario, compared to \avfuzzer. These results demonstrate that \Legend is more effective and efficient in exposing more diverse critical scenarios.

While \Legend has additional time cost during the two-phase conversion process, such as the API calls to LLMs, we observe that its overall time cost is lower than that of the two baseline techniques, averaging 32.1 seconds per scenario. This efficiency can be attributed to two main factors: (1) the time required for API calls is minimal, compared to the duration of thousands of simulations;
(2) as \Legend pursues diversity, the test cases executed by \Legend are often more lightweight than the cases executed by \avfuzzer that tries to search for more complicated cases for criticality. Usually, simpler test cases can also expose more serious system defects, since they are not supposed to be there.
These factors contribute to the reduced time cost, enhancing the overall efficiency of \Legend compared to the baseline techniques.

\begin{table}[!tb]
\resizebox{0.98\linewidth}{!}{
  \begin{threeparttable}
    \caption{The comparison with baselines}
    \label{tab:comparison_with_baselines}
    
     \begin{tabular}{ccccc}
     \toprule
     \multicolumn{2}{c}{\multirow{2}{*}{\textbf{Metric}}} & \multicolumn{3}{c}{\textbf{Method}} \\
     \cmidrule(l){3-5}
     ~ & ~ & \textbf{\Legend}  & \textbf{\random}  & \textbf{\avfuzzer}  \\
     \midrule
     \multicolumn{2}{c}{$\textsf{\#Types}$} & 11 & 2 & 4 \\
     \hline

     \multirow{2}{*}{$\textsf{\#TypeExposRate}$} & avg & $11.5\%$ & $3.1\%$ & $1.3\%$ \\
     ~ & std & $1.6\%$ & $0.4\%$ & $0.2\%$ \\
     \hline
     
    \multirow{2}{*}{$\textsf{\#SimForFirstType}$} & avg & 11 & 35 & 14 \\
    ~ & std & 5.2 & 17.7 & 6.7 \\
    \hline
     
     \multirow{2}{*}{$\textsf{\#SimForAllTypes}$} & avg & 1354 & 642 & 594 \\
     ~ & std & 37.5 & 145.1 & 71.2 \\
     \hline

      \multirow{2}{*}{$\textsf{\#TimeForOneScenario}$} & avg & 32.1 & 46.2 & 46.6 \\
      ~ & std & 1.5 & 0.3 & 0.5 \\
      
     \bottomrule
     \end{tabular}
     \begin{tablenotes}
      \small
      \item The terms `avg' and `std' refer to the average value and standard deviation, respectively.
    \end{tablenotes}
  \end{threeparttable}
  }
\end{table}

\begin{tcolorbox}[colback=white,colframe=black, arc=.3em, boxsep=-0.5mm]
\textbf{Answer to RQ1:} \textbf{Compared with existing techniques, \Legend demonstrates an advantage in detecting  critical scenarios with high diversity. }
\end{tcolorbox}

\subsection{RQ2: Can interactive pattern sequences help to improve transformation precision?}
\label{subsec:rq2}
To evaluate the effectiveness of the intermediate representation in our approach, we also implement a variant version of $\Legend$, called \legendminus, in which the logical scenarios are generated directly from the accident report using a single LLM.
We run $\Legend$ and \legendminus for 1400 simulations respectively on the same set of accident reports. The comparison results are presented in Table~\ref{tab:comparison_with_variant}. During the execution, \legendminus identifies 28 critical scenarios revealing potential system faults, with the first collision occurring in the 81-$st$ simulation. These critical scenarios are categorized into 2 distinct types, achieving a type exposure rate of $7.1\%$. Both types were discovered within the first 982 simulations.
In comparison, \Legend is able to generate more critical scenarios and expose more diverse types of scenarios. Notably, all types of critical scenarios discovered by \legendminus are also detected by \Legend in the 118-$th$ and 938-$th$ simulations. This demonstrates that \Legend not only identifies more critical scenarios but also uncovers a broader spectrum of distinct types, compared to \legendminus.

\begin{table}[!tb]
\resizebox{0.90\linewidth}{!}{
  \begin{threeparttable}
    \caption{The comparison between  $\Legend$ and \legendminus}
    \label{tab:comparison_with_variant}
    
     \begin{tabular}{cccc}
     \toprule
     \multicolumn{2}{c}{\multirow{2}{*}{\textbf{Metric}}} & \multicolumn{2}{c}{\textbf{Method}} \\
     \cmidrule(l){3-4}
     ~ & ~ & \textbf{\Legend}  & \textbf{\legendminus} \\
     \midrule
     \multicolumn{2}{c}{$\textsf{\#Types}$} & 11 & 2 \\
     \hline

     \multirow{2}{*}{$\textsf{\#TypeExposRate}$} & avg & $11.5\%$ & $7.1\%$ \\
     ~ & std & $1.6\%$ & $0.9\%$\\
     \hline
     
    \multirow{2}{*}{$\textsf{\#SimForFirstType}$} & avg & 11 & 81 \\
    ~ & std & 5.2 & 13.1 \\
    \hline
     
     \multirow{2}{*}{$\textsf{\#SimForAllTypes}$} & avg & 1354 & 982 \\
     ~ & std & 37.5 & 84.3 \\
     \hline

      \multirow{2}{*}{$\textsf{\#TimeForOneScenario}$} & avg & 32.1 & 31.9 \\
      ~ & std & 1.5 & 0.9 \\
      
     \bottomrule
     \end{tabular}
     \begin{tablenotes}
     \footnotesize
      \item The terms `avg' and `std' refer to the average value and standard deviation, respectively.
    \end{tablenotes}
  \end{threeparttable}
  }
\end{table}

\begin{table}[!tb]
\resizebox{0.90\linewidth}{!}{
  \begin{threeparttable}
    \caption{Cause Distribution}
    \label{tab:cause_distribution}
    
     \centering
    \begin{tabular}{p{0.12\linewidth} >{\centering\arraybackslash}p{0.1\linewidth} >{\centering\arraybackslash}p{0.1\linewidth}
    >{\centering\arraybackslash}p{0.1\linewidth}
    >{\centering\arraybackslash}p{0.1\linewidth}
     >{\centering\arraybackslash}p{0.1\linewidth}}
    \toprule
     $\textbf{\#}$ & $\textbf{C1}$ & $\textbf{C2}$ & $\textbf{C3}$ & $\textbf{C4}$ & \textbf{Total}
    \\
    \midrule
    Number & $1$ & $4$ & $2$ & $2$ & 9 \\
    \bottomrule
    \end{tabular}
  \end{threeparttable}
  }
\end{table}

To investigate why \legendminus exposes fewer types of critical scenarios than \Legend, we further conduct a detailed comparison of the logical scenarios generated by the two approaches. Specifically, we focus on the accident reports that lead to the identification of a distinct type of critical scenarios by \Legend but not by \legendminus. These accident reports, along with the logical scenarios generated by both approaches are then assigned to three authors of this paper as the assessors for the evaluation. The assessors are required to complete the following three stages for each accident report: (1) thoroughly read the original report and mark the necessary information related to the crash, such as interactions between vehicles; (2) compare the differences between the two logical scenarios represented in the same DSL; (3) record the main reasons for the differences in experimental results.

Finally, the assessment results are consolidated in a consensus meeting where the assessors compare their decisions. Any conflicting decisions are resolved through discussion. As a result, we conclude the following root causes:
\begin{compactitem}[$\bullet$]
    \item The logical scenario has an incorrect trigger sequence for the maneuvers of NPC vehicles, represented by $\textbf{C1}$;
    
    \item The logical scenario lacks the necessary maneuvers of NPC vehicles that facilitate a violation, represented by $\textbf{C2}$;
    
    \item The logical scenario includes redundant maneuvers that could affect the substitution of the ego vehicle, represented by $\textbf{C3}$;
    
    \item The logical scenario contains parameter ranges that could not accurately reflect the interactions indicated from the original report, represented by $\textbf{C4}$.
\end{compactitem}
\smallskip
The percentage of each type of cause is listed in Table~\ref{tab:cause_distribution}. It can be observed that $\textbf{C4}$ is the main cause for 44.4\% (4 of 9) of the cases. Moreover, $\textbf{C1}$, $\textbf{C2}$ and $\textbf{C4}$ have a direct relation to the interactive actions between vehicles described in the original accident reports, accounting for 77.8\% (7 of 9) of the causes. These findings emphasize the necessity of a semi-formal functional scenario, particularly the extraction of interactive pattern sequences, in improving the performance of \Legend. 

\begin{tcolorbox}[colback=white,colframe=black, arc=.3em, boxsep=-0.5mm]
\textbf{Answer to RQ2:} \textbf{The two-phase conversion process is essential. The intermediate representation plays a crucial role in ensuring transformation precision and exposing more diverse critical scenarios.}
\end{tcolorbox}

\subsection{RQ3: How accurate are \llmone and \llmtwo for the transformation tasks in \Legend?}
\label{subsec:rq3}
As mentioned in \S{}\ref{sec:experiment_design}, we address RQ3 through a user study to investigate the accuracy of the two-stage conversion performed by LLMs. Table~\ref{tab:user_study} lists the statements and the selected accident reports utilized in the study. We construct five statements specifically for evaluation purposes. The statements related to the intermediate representation cover aspects such as road structures (Statement \#1), initial vehicle actions (Statement \#2), and the interactive patterns (Statement \#3). The statements related to logical scenarios focus on the correctness of the logical scenario template (Statement \#4) and the parameter ranges (Statement \#5).
For the evaluation, we select 5 accident reports, with the main distinguishing features summarized in Table~\ref{tab:user_study}. The selected reports cover a wide range of road structures, numbers of vehicles, and crash types, which facilitate a comprehensive assessment of the two-stage conversion process in \Legend.

\begin{table}[!tb]
  \centering
  \caption{Statements and report cases used for the user study}
  \label{tab:user_study} 
  \renewcommand{\arraystretch}{1.2}{
  \begin{subtable}[t]{\linewidth}
    \resizebox{\linewidth}{!}{
    \small
    \begin{tabular}{p{0.03\linewidth} p{0.97\linewidth}}
    \toprule
  \textbf{ID} & \textbf{Statement} 
    \\
    \midrule
      1 & The description of the road structure is correct. \\     
      2 & The initial actions of vehicles are described accurately. \\
      3 & The interactive patterns between vehicles are extracted accurately. \\
     \midrule
      4 & The formal test case template matches the interactive pattern sequence. \\     
      5 & The parameter ranges in the test case can accurately reflect the scenario. \\
    \bottomrule
    \end{tabular}}
    \end{subtable}
    
 \hfill

 \begin{subtable}[t]{\linewidth}
    \resizebox{\linewidth}{!}{
    \begin{tabular}{p{0.1\linewidth}  p{0.25\linewidth} p{0.65\linewidth}}
    \toprule
    \textbf{\#} & \textbf{Report ID} & \textbf{Short Description} 
    \\
    \midrule
     Case 1 & 2005008586482 & 3-lane straight road, 3 vehicles involved, rear-end collision \\
     
     Case 2 & 2006009501304 & 4-lane curved road, 2 vehicles involved, rear-end collision \\ 
     
     Case 3 & 2005004496082 & 4-lane straight road, 3 vehicles involved, lane-change collision \\
     
     Case 4 & 2006002585306 & 4-lane curved road, 2 vehicles involved, lane-change collision \\
     
     Case 5 & 2006043699404 & 6-lane straight road, 3 vehicles involved, lane-change collision \\
    
    \bottomrule
    \end{tabular}}
    \end{subtable}
 }
\end{table}

\begin{figure}[!tb]
    \centering
    \includegraphics[width=0.9\linewidth, scale=0.5]{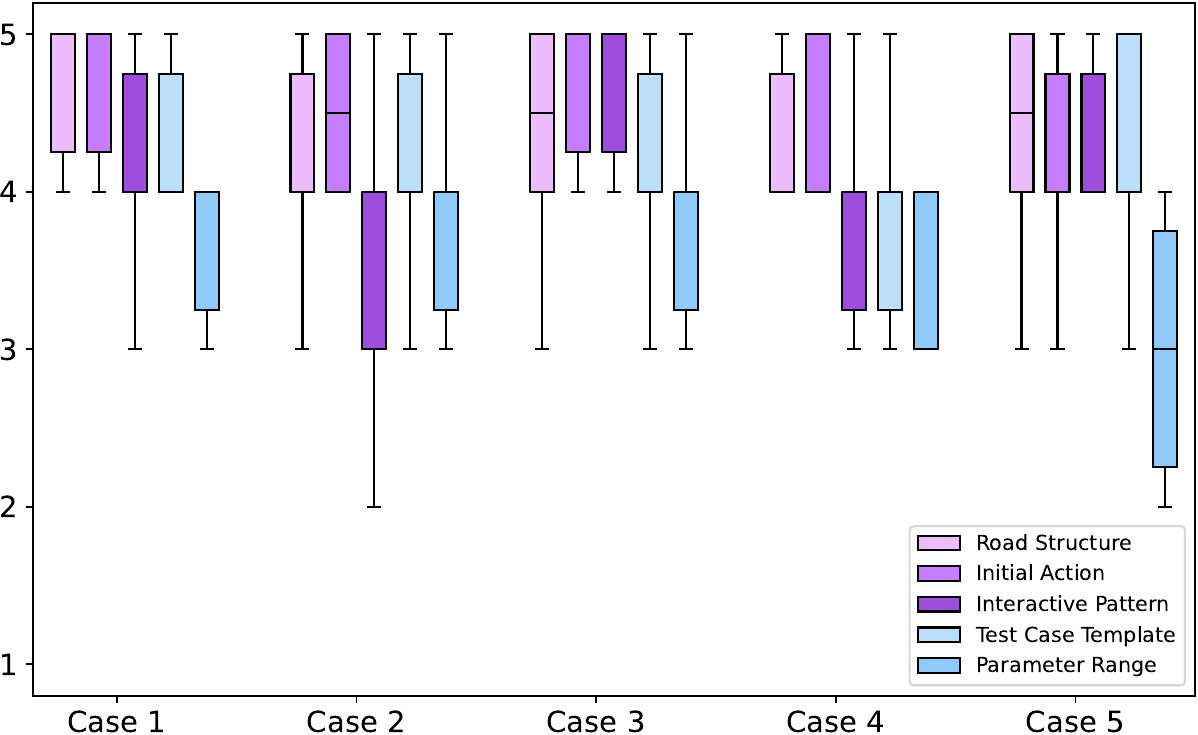}
    \caption{Results over different cases}
    \label{fig:use_study1}
\end{figure}

Fig.~\ref{fig:use_study1} presents the results of participants' responses across the five cases. The data indicates that participants generally agree that \Legend accurately extracts the road structures and initial vehicle actions. Regarding the extracted interactive pattern sequences across the five cases, participants still agree but less strongly for $Case 2$ and $Case 4$. By the analysis of their feedback, we find that this discrepancy arises because the large language model (LLM) can generate new actions for vehicles when specific actions are not explicitly described in the accident report. This result aligns with the design choice of \Legend, as LLMs are generative models characterized by their high creativity. Moreover, while the template conversion process is deemed accurate by participants, the parameter filling process is found to be less precise. This is attributed to the current limitations in the reasoning capabilities of LLMs.

\begin{tcolorbox}[colback=white,colframe=black, arc=.3em, boxsep=-0.5mm]
\textbf{Answer to RQ3:} \textbf{Overall, the transformations by LLMs are accurate. Although LLMs currently exhibit limitations in parameter filling for test cases, \Legend mitigates this issue by adopting only a part of the parameters that have higher confidence.}
\end{tcolorbox}

\subsection{Threats to Validity}
\myparagraph{Internal Validity} The threat to internal validity could be the possibility of instrumentation effects stemming from faults within our approach \Legend. These effects could introduce unintended biases or inaccuracies in the generated scenarios and evaluation results. To mitigate these threats, we conducted thorough testing of the implementation and inspected the intermediate results during execution.
To avoid bias in the user survey, we anonymized it and assigned tasks to participants randomly.

\myparagraph{External Validity} The threats to external validity could be the selected accident reports. We only selected a small number of police reports from the NHTSA database to conduct the experiments; however, we ensured that the selected reports cover different crash types, road geometries, number of cars, and car movements. Since there exists a perception delay in the software simulator, i.e., LGSVL, we mitigate this by bypassing the perception module and sending the ground truth of the perception to the ADS. Another threat is the generalizability of our approach to other types of LLMs and ADSs under test. To mitigate this threat, we selected GPT-4~\cite{achiam2023gpt} for the extraction and conversion tasks, as it was one of the most advanced and powerful LLMs available as of the time of performing our experiments. For the ADS under test, we adopted Baidu Apollo~\cite{apollo}, an open-source, industrial-grade ADS that is one of the most popular platforms and has been extensively studied~\cite{li2020av, cheng2023behavexplor, huai2023doppelganger, zhong2022neural} in the literature of autonomous driving. Therefore, we believe the selected LLM and ADS under test are representative for our approach. A final threat to external validity is the eligibility of the participants of our user study. 
To mitigate this issue, we involved graduate students who have been working extensively in this domain for multiple years, and moreover, these participants were required to  read the accident reports thoroughly and  rate their agreement with the corresponding questions carefully, therefore, they are knowledgeable about the problem under investigation and able to make precise and reliable assessments. 

\section{Related Work}
\label{sec:related_work}
\myparagraph{Large Language Models} have been widely adopted due to their extraordinary performances in different downstream tasks~\cite{min2023recent, lemieux2023codamosa, pan2023understanding}.
Currently, there are two main techniques that employ LLMs for domain-specific tasks, namely, prompt engineering~\cite{liu2023pre} and fine-tuning~\cite{radford2018improving}. Prompt engineering, such as few-shot learning~\cite{brown2020language}, works by providing a description of the task  with a few demonstrations to directly elicit the desired outputs. In contrast, fine-tuning involves training a pre-trained model on a task-specific dataset to adapt it to particular tasks, e.g., code generation~\cite{roziere2023code}. However, since fine-tuning an LLM is time-consuming and expensive, prompting strategies have been increasingly adopted~\cite{zhang2023multimodal}.
In this work, our LLM-related tasks are accomplished via prompt engineering.

\myparagraph{Module-level ADS testing} targets the evaluation of individual components within an ADS, including the perception, planning, and control modules. Testing of the perception module involves generating adversarial inputs to challenge the machine learning models dedicated to various perception tasks, such as object detection~\cite{sun2020towards, cao2021invisible}, semantic segmentation~\cite{xu2020adversarial}, and traffic sign recognition~\cite{li2020adaptive}. For example, Cao et al.~\cite{cao2021invisible} propose an optimization-based attack method aimed at generating adversarial 3D objects to deceive both camera-based and LiDAR-based object detectors within ADS perception systems.
Meanwhile, testing of the planning and control modules primarily focuses on assessing the safety of planned trajectories~\cite{laurent2019mutation, laurent2020achieving} or verifying the correctness of control decision~\cite{djoudi2020simulation}. These evaluations often rely on dedicated path planners and controllers~\cite{laurent2019mutation, laurent2020achieving, djoudi2020simulation}.
However, existing module-level testing approaches tend to overlook the potential issues arising from the complex collaborations among different modules.

\myparagraph{System-level ADS testing} considers the interactions and dependencies among its different modules, and often evaluates an ADS by system simulation.
To effectively find critical scenarios that can reveal the system faults, various approaches have been proposed including search-based testing~\cite{gambi2019automatically, riccio2020model, li2020av, tian2022mosat, zhong2022neural,  cheng2023behavexplor, huai2023doppelganger, tang2023evoscenario, zhou2023collision,zohdinasab2021deephyperion}, metamorphic testing~\cite{zhou2019metamorphic, han2020metamorphic} and formal methods~\cite{sun2022lawbreaker, zhang2023testing, zhou2023specification}. Among these methodologies, search-based techniques are prominent for detecting critical scenarios in a huge input space. For instance, Tian et al.~\cite{tian2022mosat} encode atomic driving behaviors as motif patterns, and apply a multi-objective genetic algorithm to generate diverse test scenarios. Zhong et.al~\cite{zhong2022neural} employ a neural network to guide the process of seed selection, thereby accelerating evolutionary search.
There is a line of work~\cite{sippl2018distributed, gambi2019generating, zhang2023building} that aim to enhance realism of test cases. For example, Gambi et al.~\cite{gambi2019generating} propose \emph{AC3R}, which can recover vehicle collisions from public accident reports based on natural language processing techniques. Zhang et al.~\cite{zhang2023building} employ a segmentation model to extract information for reconstructing scenes derived from actual accident data. 
However, most of these studies~\cite{li2020av, tian2022mosat, huai2023doppelganger, tang2023evoscenario, zhou2023collision, sun2022lawbreaker, zhang2023testing, zhou2023specification, gambi2019generating, zhang2023building}  focus on generating or reconstructing concrete scenarios without considering functional levels, so they cannot control scenario diversity at a high level.

\section{Conclusion and Future Work}
\label{sec:conclusion}
Detecting diverse safety violations is crucial for ADS testing, because it can expose system defects in different aspects and avoid redundant search space exploration. In this paper, we propose \Legend that features a top-down style of scenario generation for testing of the ADS that starts with functional scenarios and then steps downwards. Specifically, \Legend employs two LLMs to transform  functional scenarios to formal logical scenarios, and then searches with the logical scenarios for critical scenarios.
As future work, we plan to try functional scenarios from other sources to further study the performances of LLMs in ADS testing. 

\begin{acks}
This work was supported in part by the National Key R\&D Program of China (2023YFF0612300), in part by the Anhui Provincial Department of Science and Technology under Grant 202103a05020009.
Z. Zhang is supported by JSPS KAKENHI Grant No. JP23K16865 and No. JP23H03372.
\end{acks}

\bibliographystyle{ACM-Reference-Format}
\bibliography{ACM}


\begin{thebibliography}{50}


\ifx \showCODEN    \undefined \def \showCODEN     #1{\unskip}     \fi
\ifx \showDOI      \undefined \def \showDOI       #1{#1}\fi
\ifx \showISBNx    \undefined \def \showISBNx     #1{\unskip}     \fi
\ifx \showISBNxiii \undefined \def \showISBNxiii  #1{\unskip}     \fi
\ifx \showISSN     \undefined \def \showISSN      #1{\unskip}     \fi
\ifx \showLCCN     \undefined \def \showLCCN      #1{\unskip}     \fi
\ifx \shownote     \undefined \def \shownote      #1{#1}          \fi
\ifx \showarticletitle \undefined \def \showarticletitle #1{#1}   \fi
\ifx \showURL      \undefined \def \showURL       {\relax}        \fi
\providecommand\bibfield[2]{#2}
\providecommand\bibinfo[2]{#2}
\providecommand\natexlab[1]{#1}
\providecommand\showeprint[2][]{arXiv:#2}

\bibitem[gpt(2023)]%
        {gpt4endpoint}
 \bibinfo{year}{2023}\natexlab{}.
\newblock \bibinfo{title}{Models - GPT-4}.
\newblock \bibinfo{howpublished}{\url{https://platform.openai.com/docs/models/gpt- 4}}.
\newblock


\bibitem[Abuelenin and Abul-Magd(2015)]%
        {abuelenin2015effect}
\bibfield{author}{\bibinfo{person}{Sherif~M Abuelenin} {and} \bibinfo{person}{Adel~Y Abul-Magd}.} \bibinfo{year}{2015}\natexlab{}.
\newblock \showarticletitle{Effect of minimum headway distance on connectivity of VANETs}.
\newblock \bibinfo{journal}{\emph{AEU-International Journal of Electronics and Communications}} \bibinfo{volume}{69}, \bibinfo{number}{5} (\bibinfo{year}{2015}), \bibinfo{pages}{867--871}.
\newblock


\bibitem[Achiam et~al\mbox{.}(2023)]%
        {achiam2023gpt}
\bibfield{author}{\bibinfo{person}{Josh Achiam}, \bibinfo{person}{Steven Adler}, \bibinfo{person}{Sandhini Agarwal}, \bibinfo{person}{Lama Ahmad}, \bibinfo{person}{Ilge Akkaya}, \bibinfo{person}{Florencia~Leoni Aleman}, \bibinfo{person}{Diogo Almeida}, \bibinfo{person}{Janko Altenschmidt}, \bibinfo{person}{Sam Altman}, \bibinfo{person}{Shyamal Anadkat}, {et~al\mbox{.}}} \bibinfo{year}{2023}\natexlab{}.
\newblock \showarticletitle{Gpt-4 technical report}.
\newblock \bibinfo{journal}{\emph{arXiv preprint arXiv:2303.08774}} (\bibinfo{year}{2023}).
\newblock


\bibitem[Baidu({[n.\,d.]})]%
        {apollo}
\bibfield{author}{\bibinfo{person}{Baidu}.} \bibinfo{year}{[n.\,d.]}\natexlab{}.
\newblock \bibinfo{title}{An open autonomous driving platform}.
\newblock \bibinfo{howpublished}{\url{https://github.com/ApolloAuto/apollo}}.
\newblock


\bibitem[Brown et~al\mbox{.}(2020)]%
        {brown2020language}
\bibfield{author}{\bibinfo{person}{Tom Brown}, \bibinfo{person}{Benjamin Mann}, \bibinfo{person}{Nick Ryder}, \bibinfo{person}{Melanie Subbiah}, \bibinfo{person}{Jared~D Kaplan}, \bibinfo{person}{Prafulla Dhariwal}, \bibinfo{person}{Arvind Neelakantan}, \bibinfo{person}{Pranav Shyam}, \bibinfo{person}{Girish Sastry}, \bibinfo{person}{Amanda Askell}, {et~al\mbox{.}}} \bibinfo{year}{2020}\natexlab{}.
\newblock \showarticletitle{Language models are few-shot learners}.
\newblock \bibinfo{journal}{\emph{Advances in neural information processing systems}}  \bibinfo{volume}{33} (\bibinfo{year}{2020}), \bibinfo{pages}{1877--1901}.
\newblock


\bibitem[Cal{\`o} et~al\mbox{.}(2020)]%
        {calo2020generating}
\bibfield{author}{\bibinfo{person}{Alessandro Cal{\`o}}, \bibinfo{person}{Paolo Arcaini}, \bibinfo{person}{Shaukat Ali}, \bibinfo{person}{Florian Hauer}, {and} \bibinfo{person}{Fuyuki Ishikawa}.} \bibinfo{year}{2020}\natexlab{}.
\newblock \showarticletitle{Generating avoidable collision scenarios for testing autonomous driving systems}. In \bibinfo{booktitle}{\emph{2020 IEEE 13th International Conference on Software Testing, Validation and Verification (ICST)}}. IEEE, \bibinfo{pages}{375--386}.
\newblock


\bibitem[Cao et~al\mbox{.}(2021)]%
        {cao2021invisible}
\bibfield{author}{\bibinfo{person}{Yulong Cao}, \bibinfo{person}{Ningfei Wang}, \bibinfo{person}{Chaowei Xiao}, \bibinfo{person}{Dawei Yang}, \bibinfo{person}{Jin Fang}, \bibinfo{person}{Ruigang Yang}, \bibinfo{person}{Qi~Alfred Chen}, \bibinfo{person}{Mingyan Liu}, {and} \bibinfo{person}{Bo Li}.} \bibinfo{year}{2021}\natexlab{}.
\newblock \showarticletitle{Invisible for both camera and lidar: Security of multi-sensor fusion based perception in autonomous driving under physical-world attacks}. In \bibinfo{booktitle}{\emph{2021 IEEE Symposium on Security and Privacy (SP)}}. IEEE, \bibinfo{pages}{176--194}.
\newblock


\bibitem[Cheng et~al\mbox{.}(2023)]%
        {cheng2023behavexplor}
\bibfield{author}{\bibinfo{person}{Mingfei Cheng}, \bibinfo{person}{Yuan Zhou}, {and} \bibinfo{person}{Xiaofei Xie}.} \bibinfo{year}{2023}\natexlab{}.
\newblock \showarticletitle{BehAVExplor: Behavior Diversity Guided Testing for Autonomous Driving Systems}. In \bibinfo{booktitle}{\emph{Proceedings of the 32nd ACM SIGSOFT International Symposium on Software Testing and Analysis}}. \bibinfo{pages}{488--500}.
\newblock


\bibitem[Deaths({[n.\,d.]})]%
        {tesla_death}
\bibfield{author}{\bibinfo{person}{Tesla Deaths}.} \bibinfo{year}{[n.\,d.]}\natexlab{}.
\newblock \bibinfo{title}{A record of Tesla-related fatalities}.
\newblock \bibinfo{howpublished}{\url{https://www.tesladeaths.com/}}.
\newblock


\bibitem[Deb et~al\mbox{.}(2002)]%
        {deb2002fast}
\bibfield{author}{\bibinfo{person}{Kalyanmoy Deb}, \bibinfo{person}{Amrit Pratap}, \bibinfo{person}{Sameer Agarwal}, {and} \bibinfo{person}{TAMT Meyarivan}.} \bibinfo{year}{2002}\natexlab{}.
\newblock \showarticletitle{A fast and elitist multiobjective genetic algorithm: NSGA-II}.
\newblock \bibinfo{journal}{\emph{IEEE transactions on evolutionary computation}} \bibinfo{volume}{6}, \bibinfo{number}{2} (\bibinfo{year}{2002}), \bibinfo{pages}{182--197}.
\newblock


\bibitem[Djoudi et~al\mbox{.}(2020)]%
        {djoudi2020simulation}
\bibfield{author}{\bibinfo{person}{Adel Djoudi}, \bibinfo{person}{Loic Coquelin}, {and} \bibinfo{person}{R{\'e}mi Regnier}.} \bibinfo{year}{2020}\natexlab{}.
\newblock \showarticletitle{A simulation-based framework for functional testing of automated driving controllers}. In \bibinfo{booktitle}{\emph{2020 IEEE 23rd International Conference on Intelligent Transportation Systems (ITSC)}}. IEEE, \bibinfo{pages}{1--6}.
\newblock


\bibitem[Gambi et~al\mbox{.}(2019a)]%
        {gambi2019generating}
\bibfield{author}{\bibinfo{person}{Alessio Gambi}, \bibinfo{person}{Tri Huynh}, {and} \bibinfo{person}{Gordon Fraser}.} \bibinfo{year}{2019}\natexlab{a}.
\newblock \showarticletitle{Generating effective test cases for self-driving cars from police reports}. In \bibinfo{booktitle}{\emph{Proceedings of the 2019 27th ACM Joint Meeting on European Software Engineering Conference and Symposium on the Foundations of Software Engineering}}. \bibinfo{pages}{257--267}.
\newblock


\bibitem[Gambi et~al\mbox{.}(2019b)]%
        {gambi2019automatically}
\bibfield{author}{\bibinfo{person}{Alessio Gambi}, \bibinfo{person}{Marc Mueller}, {and} \bibinfo{person}{Gordon Fraser}.} \bibinfo{year}{2019}\natexlab{b}.
\newblock \showarticletitle{Automatically testing self-driving cars with search-based procedural content generation}. In \bibinfo{booktitle}{\emph{Proceedings of the 28th ACM SIGSOFT International Symposium on Software Testing and Analysis}}. \bibinfo{pages}{318--328}.
\newblock


\bibitem[Han and Zhou(2020)]%
        {han2020metamorphic}
\bibfield{author}{\bibinfo{person}{Jia~Cheng Han} {and} \bibinfo{person}{Zhi~Quan Zhou}.} \bibinfo{year}{2020}\natexlab{}.
\newblock \showarticletitle{Metamorphic fuzz testing of autonomous vehicles}. In \bibinfo{booktitle}{\emph{Proceedings of the IEEE/ACM 42nd International Conference on Software Engineering Workshops}}. \bibinfo{pages}{380--385}.
\newblock


\bibitem[Huai et~al\mbox{.}(2023)]%
        {huai2023doppelganger}
\bibfield{author}{\bibinfo{person}{Yuqi Huai}, \bibinfo{person}{Yuntianyi Chen}, \bibinfo{person}{Sumaya Almanee}, \bibinfo{person}{Tuan Ngo}, \bibinfo{person}{Xiang Liao}, \bibinfo{person}{Ziwen Wan}, \bibinfo{person}{Qi~Alfred Chen}, {and} \bibinfo{person}{Joshua Garcia}.} \bibinfo{year}{2023}\natexlab{}.
\newblock \showarticletitle{Doppelg{\"a}nger test generation for revealing bugs in autonomous driving software}. In \bibinfo{booktitle}{\emph{2023 IEEE/ACM 45th International Conference on Software Engineering (ICSE)}}. IEEE, \bibinfo{pages}{2591--2603}.
\newblock


\bibitem[Lande et~al\mbox{.}(1987)]%
        {lande1987effective}
\bibfield{author}{\bibinfo{person}{Russell Lande}, \bibinfo{person}{George~F Barrowclough}, {et~al\mbox{.}}} \bibinfo{year}{1987}\natexlab{}.
\newblock \showarticletitle{Effective population size, genetic variation, and their use in population management}.
\newblock \bibinfo{journal}{\emph{Viable populations for conservation}}  \bibinfo{volume}{87} (\bibinfo{year}{1987}), \bibinfo{pages}{87--124}.
\newblock


\bibitem[Laurent et~al\mbox{.}(2019)]%
        {laurent2019mutation}
\bibfield{author}{\bibinfo{person}{Thomas Laurent}, \bibinfo{person}{Paolo Arcaini}, \bibinfo{person}{Fuyuki Ishikawa}, {and} \bibinfo{person}{Anthony Ventresque}.} \bibinfo{year}{2019}\natexlab{}.
\newblock \showarticletitle{A mutation-based approach for assessing weight coverage of a path planner}. In \bibinfo{booktitle}{\emph{2019 26th Asia-Pacific Software Engineering Conference (APSEC)}}. IEEE, \bibinfo{pages}{94--101}.
\newblock


\bibitem[Laurent et~al\mbox{.}(2020)]%
        {laurent2020achieving}
\bibfield{author}{\bibinfo{person}{Thomas Laurent}, \bibinfo{person}{Paolo Arcaini}, \bibinfo{person}{Fuyuki Ishikawa}, {and} \bibinfo{person}{Anthony Ventresque}.} \bibinfo{year}{2020}\natexlab{}.
\newblock \showarticletitle{Achieving Weight Coverage for an Autonomous Driving System with Search-based Test Generation}. In \bibinfo{booktitle}{\emph{2020 25th International Conference on Engineering of Complex Computer Systems (ICECCS)}}. IEEE, \bibinfo{pages}{93--102}.
\newblock


\bibitem[Lemieux et~al\mbox{.}(2023)]%
        {lemieux2023codamosa}
\bibfield{author}{\bibinfo{person}{Caroline Lemieux}, \bibinfo{person}{Jeevana~Priya Inala}, \bibinfo{person}{Shuvendu~K Lahiri}, {and} \bibinfo{person}{Siddhartha Sen}.} \bibinfo{year}{2023}\natexlab{}.
\newblock \showarticletitle{Codamosa: Escaping coverage plateaus in test generation with pre-trained large language models}. In \bibinfo{booktitle}{\emph{2023 IEEE/ACM 45th International Conference on Software Engineering (ICSE)}}. IEEE, \bibinfo{pages}{919--931}.
\newblock


\bibitem[Li et~al\mbox{.}(2020a)]%
        {li2020av}
\bibfield{author}{\bibinfo{person}{Guanpeng Li}, \bibinfo{person}{Yiran Li}, \bibinfo{person}{Saurabh Jha}, \bibinfo{person}{Timothy Tsai}, \bibinfo{person}{Michael Sullivan}, \bibinfo{person}{Siva Kumar~Sastry Hari}, \bibinfo{person}{Zbigniew Kalbarczyk}, {and} \bibinfo{person}{Ravishankar Iyer}.} \bibinfo{year}{2020}\natexlab{a}.
\newblock \showarticletitle{Av-fuzzer: Finding safety violations in autonomous driving systems}. In \bibinfo{booktitle}{\emph{2020 IEEE 31st International Symposium on Software Reliability Engineering (ISSRE)}}. IEEE, \bibinfo{pages}{25--36}.
\newblock


\bibitem[Li et~al\mbox{.}(2020b)]%
        {li2020adaptive}
\bibfield{author}{\bibinfo{person}{Yujie Li}, \bibinfo{person}{Xing Xu}, \bibinfo{person}{Jinhui Xiao}, \bibinfo{person}{Siyuan Li}, {and} \bibinfo{person}{Heng~Tao Shen}.} \bibinfo{year}{2020}\natexlab{b}.
\newblock \showarticletitle{Adaptive square attack: Fooling autonomous cars with adversarial traffic signs}.
\newblock \bibinfo{journal}{\emph{IEEE Internet of Things Journal}} \bibinfo{volume}{8}, \bibinfo{number}{8} (\bibinfo{year}{2020}), \bibinfo{pages}{6337--6347}.
\newblock


\bibitem[Likert(1932)]%
        {likert1932technique}
\bibfield{author}{\bibinfo{person}{Rensis Likert}.} \bibinfo{year}{1932}\natexlab{}.
\newblock \showarticletitle{A technique for the measurement of attitudes.}
\newblock \bibinfo{journal}{\emph{Archives of psychology}} (\bibinfo{year}{1932}).
\newblock


\bibitem[Liu et~al\mbox{.}(2023)]%
        {liu2023pre}
\bibfield{author}{\bibinfo{person}{Pengfei Liu}, \bibinfo{person}{Weizhe Yuan}, \bibinfo{person}{Jinlan Fu}, \bibinfo{person}{Zhengbao Jiang}, \bibinfo{person}{Hiroaki Hayashi}, {and} \bibinfo{person}{Graham Neubig}.} \bibinfo{year}{2023}\natexlab{}.
\newblock \showarticletitle{Pre-train, prompt, and predict: A systematic survey of prompting methods in natural language processing}.
\newblock \bibinfo{journal}{\emph{Comput. Surveys}} \bibinfo{volume}{55}, \bibinfo{number}{9} (\bibinfo{year}{2023}), \bibinfo{pages}{1--35}.
\newblock


\bibitem[Menzel et~al\mbox{.}(2019)]%
        {menzel2019functional}
\bibfield{author}{\bibinfo{person}{Till Menzel}, \bibinfo{person}{Gerrit Bagschik}, \bibinfo{person}{Leon Isensee}, \bibinfo{person}{Andre Schomburg}, {and} \bibinfo{person}{Markus Maurer}.} \bibinfo{year}{2019}\natexlab{}.
\newblock \showarticletitle{From functional to logical scenarios: Detailing a keyword-based scenario description for execution in a simulation environment}. In \bibinfo{booktitle}{\emph{2019 IEEE Intelligent Vehicles Symposium (IV)}}. IEEE, \bibinfo{pages}{2383--2390}.
\newblock


\bibitem[Menzel et~al\mbox{.}(2018)]%
        {menzel2018scenarios}
\bibfield{author}{\bibinfo{person}{Till Menzel}, \bibinfo{person}{Gerrit Bagschik}, {and} \bibinfo{person}{Markus Maurer}.} \bibinfo{year}{2018}\natexlab{}.
\newblock \showarticletitle{Scenarios for development, test and validation of automated vehicles}. In \bibinfo{booktitle}{\emph{2018 IEEE Intelligent Vehicles Symposium (IV)}}. IEEE, \bibinfo{pages}{1821--1827}.
\newblock


\bibitem[Min et~al\mbox{.}(2023)]%
        {min2023recent}
\bibfield{author}{\bibinfo{person}{Bonan Min}, \bibinfo{person}{Hayley Ross}, \bibinfo{person}{Elior Sulem}, \bibinfo{person}{Amir Pouran~Ben Veyseh}, \bibinfo{person}{Thien~Huu Nguyen}, \bibinfo{person}{Oscar Sainz}, \bibinfo{person}{Eneko Agirre}, \bibinfo{person}{Ilana Heintz}, {and} \bibinfo{person}{Dan Roth}.} \bibinfo{year}{2023}\natexlab{}.
\newblock \showarticletitle{Recent advances in natural language processing via large pre-trained language models: A survey}.
\newblock \bibinfo{journal}{\emph{Comput. Surveys}} \bibinfo{volume}{56}, \bibinfo{number}{2} (\bibinfo{year}{2023}), \bibinfo{pages}{1--40}.
\newblock


\bibitem[(NHTSA)({[n.\,d.]}a)]%
        {nhtsa_case}
\bibfield{author}{\bibinfo{person}{National Highway Traffic Safety~Administration (NHTSA)}.} \bibinfo{year}{[n.\,d.]}\natexlab{a}.
\newblock \bibinfo{title}{Case ID: 2005004496082}.
\newblock \bibinfo{howpublished}{\url{https://crashviewer.nhtsa.dot.gov/nass-NMVCCS/CaseForm.aspx?xsl=main.xsl&CaseID=2005004496082}}.
\newblock


\bibitem[(NHTSA)({[n.\,d.]}b)]%
        {nhtsa}
\bibfield{author}{\bibinfo{person}{National Highway Traffic Safety~Administration (NHTSA)}.} \bibinfo{year}{[n.\,d.]}\natexlab{b}.
\newblock \bibinfo{title}{National motor vehicle crash causation survey}.
\newblock \bibinfo{howpublished}{\url{https://crashviewer.nhtsa.dot.gov/ LegacyNMVCCS/Search}}.
\newblock


\bibitem[Pan et~al\mbox{.}(2023)]%
        {pan2023understanding}
\bibfield{author}{\bibinfo{person}{Rangeet Pan}, \bibinfo{person}{Ali~Reza Ibrahimzada}, \bibinfo{person}{Rahul Krishna}, \bibinfo{person}{Divya Sankar}, \bibinfo{person}{Lambert~Pouguem Wassi}, \bibinfo{person}{Michele Merler}, \bibinfo{person}{Boris Sobolev}, \bibinfo{person}{Raju Pavuluri}, \bibinfo{person}{Saurabh Sinha}, {and} \bibinfo{person}{Reyhaneh Jabbarvand}.} \bibinfo{year}{2023}\natexlab{}.
\newblock \showarticletitle{Understanding the effectiveness of large language models in code translation}.
\newblock \bibinfo{journal}{\emph{arXiv preprint arXiv:2308.03109}} (\bibinfo{year}{2023}).
\newblock


\bibitem[Radford et~al\mbox{.}(2018)]%
        {radford2018improving}
\bibfield{author}{\bibinfo{person}{Alec Radford}, \bibinfo{person}{Karthik Narasimhan}, \bibinfo{person}{Tim Salimans}, \bibinfo{person}{Ilya Sutskever}, {et~al\mbox{.}}} \bibinfo{year}{2018}\natexlab{}.
\newblock \showarticletitle{Improving language understanding by generative pre-training}.
\newblock  (\bibinfo{year}{2018}).
\newblock


\bibitem[Riccio and Tonella(2020)]%
        {riccio2020model}
\bibfield{author}{\bibinfo{person}{Vincenzo Riccio} {and} \bibinfo{person}{Paolo Tonella}.} \bibinfo{year}{2020}\natexlab{}.
\newblock \showarticletitle{Model-based exploration of the frontier of behaviours for deep learning system testing}. In \bibinfo{booktitle}{\emph{Proceedings of the 28th ACM Joint Meeting on European Software Engineering Conference and Symposium on the Foundations of Software Engineering}}. \bibinfo{pages}{876--888}.
\newblock


\bibitem[Rong et~al\mbox{.}(2020)]%
        {rong2020lgsvl}
\bibfield{author}{\bibinfo{person}{Guodong Rong}, \bibinfo{person}{Byung~Hyun Shin}, \bibinfo{person}{Hadi Tabatabaee}, \bibinfo{person}{Qiang Lu}, \bibinfo{person}{Steve Lemke}, \bibinfo{person}{M{\=a}rti{\c{n}}{\v{s}} Mo{\v{z}}eiko}, \bibinfo{person}{Eric Boise}, \bibinfo{person}{Geehoon Uhm}, \bibinfo{person}{Mark Gerow}, \bibinfo{person}{Shalin Mehta}, {et~al\mbox{.}}} \bibinfo{year}{2020}\natexlab{}.
\newblock \showarticletitle{Lgsvl simulator: A high fidelity simulator for autonomous driving}. In \bibinfo{booktitle}{\emph{2020 IEEE 23rd International conference on intelligent transportation systems (ITSC)}}. IEEE, \bibinfo{pages}{1--6}.
\newblock


\bibitem[Roziere et~al\mbox{.}(2023)]%
        {roziere2023code}
\bibfield{author}{\bibinfo{person}{Baptiste Roziere}, \bibinfo{person}{Jonas Gehring}, \bibinfo{person}{Fabian Gloeckle}, \bibinfo{person}{Sten Sootla}, \bibinfo{person}{Itai Gat}, \bibinfo{person}{Xiaoqing~Ellen Tan}, \bibinfo{person}{Yossi Adi}, \bibinfo{person}{Jingyu Liu}, \bibinfo{person}{Tal Remez}, \bibinfo{person}{J{\'e}r{\'e}my Rapin}, {et~al\mbox{.}}} \bibinfo{year}{2023}\natexlab{}.
\newblock \showarticletitle{Code llama: Open foundation models for code}.
\newblock \bibinfo{journal}{\emph{arXiv preprint arXiv:2308.12950}} (\bibinfo{year}{2023}).
\newblock


\bibitem[Sippl et~al\mbox{.}(2018)]%
        {sippl2018distributed}
\bibfield{author}{\bibinfo{person}{Christoph Sippl}, \bibinfo{person}{Benedikt Schwab}, \bibinfo{person}{Peter Kielar}, {and} \bibinfo{person}{Anatoli Djanatliev}.} \bibinfo{year}{2018}\natexlab{}.
\newblock \showarticletitle{Distributed real-time traffic simulation for autonomous vehicle testing in urban environments}. In \bibinfo{booktitle}{\emph{2018 21st International Conference on Intelligent Transportation Systems (ITSC)}}. IEEE, \bibinfo{pages}{2562--2567}.
\newblock


\bibitem[Sun et~al\mbox{.}(2020)]%
        {sun2020towards}
\bibfield{author}{\bibinfo{person}{Jiachen Sun}, \bibinfo{person}{Yulong Cao}, \bibinfo{person}{Qi~Alfred Chen}, {and} \bibinfo{person}{Z~Morley Mao}.} \bibinfo{year}{2020}\natexlab{}.
\newblock \showarticletitle{Towards robust lidar-based perception in autonomous driving: General black-box adversarial sensor attack and countermeasures}. In \bibinfo{booktitle}{\emph{29th $\{$USENIX$\}$ Security Symposium ($\{$USENIX$\}$ Security 20)}}. \bibinfo{pages}{877--894}.
\newblock


\bibitem[Sun et~al\mbox{.}(2022)]%
        {sun2022lawbreaker}
\bibfield{author}{\bibinfo{person}{Yang Sun}, \bibinfo{person}{Christopher~M Poskitt}, \bibinfo{person}{Jun Sun}, \bibinfo{person}{Yuqi Chen}, {and} \bibinfo{person}{Zijiang Yang}.} \bibinfo{year}{2022}\natexlab{}.
\newblock \showarticletitle{LawBreaker: An approach for specifying traffic laws and fuzzing autonomous vehicles}. In \bibinfo{booktitle}{\emph{Proceedings of the 37th IEEE/ACM International Conference on Automated Software Engineering}}. \bibinfo{pages}{1--12}.
\newblock


\bibitem[Tang et~al\mbox{.}(2023a)]%
        {tang2023survey}
\bibfield{author}{\bibinfo{person}{Shuncheng Tang}, \bibinfo{person}{Zhenya Zhang}, \bibinfo{person}{Yi Zhang}, \bibinfo{person}{Jixiang Zhou}, \bibinfo{person}{Yan Guo}, \bibinfo{person}{Shuang Liu}, \bibinfo{person}{Shengjian Guo}, \bibinfo{person}{Yan-Fu Li}, \bibinfo{person}{Lei Ma}, \bibinfo{person}{Yinxing Xue}, {et~al\mbox{.}}} \bibinfo{year}{2023}\natexlab{a}.
\newblock \showarticletitle{A survey on automated driving system testing: Landscapes and trends}.
\newblock \bibinfo{journal}{\emph{ACM Transactions on Software Engineering and Methodology}} \bibinfo{volume}{32}, \bibinfo{number}{5} (\bibinfo{year}{2023}), \bibinfo{pages}{1--62}.
\newblock


\bibitem[Tang et~al\mbox{.}(2024)]%
        {replication_package}
\bibfield{author}{\bibinfo{person}{Shuncheng Tang}, \bibinfo{person}{Zhenya Zhang}, \bibinfo{person}{Jixiang Zhou}, \bibinfo{person}{Lei Lei}, \bibinfo{person}{Yuan Zhou}, {and} \bibinfo{person}{Yinxing Xue}.} \bibinfo{year}{2024}\natexlab{}.
\newblock \bibinfo{title}{LeGEND}.
\newblock \bibinfo{howpublished}{\url{https://sites.google.com/view/legend4adstesting}}.
\newblock


\bibitem[Tang et~al\mbox{.}(2023b)]%
        {tang2023evoscenario}
\bibfield{author}{\bibinfo{person}{Shuncheng Tang}, \bibinfo{person}{Zhenya Zhang}, \bibinfo{person}{Jixiang Zhou}, \bibinfo{person}{Yuan Zhou}, \bibinfo{person}{Yan-Fu Li}, {and} \bibinfo{person}{Yinxing Xue}.} \bibinfo{year}{2023}\natexlab{b}.
\newblock \showarticletitle{EvoScenario: Integrating Road Structures into Critical Scenario Generation for Autonomous Driving System Testing}. In \bibinfo{booktitle}{\emph{2023 IEEE 34th International Symposium on Software Reliability Engineering (ISSRE)}}. IEEE, \bibinfo{pages}{309--320}.
\newblock


\bibitem[Tian et~al\mbox{.}(2022)]%
        {tian2022mosat}
\bibfield{author}{\bibinfo{person}{Haoxiang Tian}, \bibinfo{person}{Yan Jiang}, \bibinfo{person}{Guoquan Wu}, \bibinfo{person}{Jiren Yan}, \bibinfo{person}{Jun Wei}, \bibinfo{person}{Wei Chen}, \bibinfo{person}{Shuo Li}, {and} \bibinfo{person}{Dan Ye}.} \bibinfo{year}{2022}\natexlab{}.
\newblock \showarticletitle{MOSAT: finding safety violations of autonomous driving systems using multi-objective genetic algorithm}. In \bibinfo{booktitle}{\emph{Proceedings of the 30th ACM Joint European Software Engineering Conference and Symposium on the Foundations of Software Engineering}}. \bibinfo{pages}{94--106}.
\newblock


\bibitem[Ulbrich et~al\mbox{.}(2015)]%
        {ulbrich2015defining}
\bibfield{author}{\bibinfo{person}{Simon Ulbrich}, \bibinfo{person}{Till Menzel}, \bibinfo{person}{Andreas Reschka}, \bibinfo{person}{Fabian Schuldt}, {and} \bibinfo{person}{Markus Maurer}.} \bibinfo{year}{2015}\natexlab{}.
\newblock \showarticletitle{Defining and substantiating the terms scene, situation, and scenario for automated driving}. In \bibinfo{booktitle}{\emph{2015 IEEE 18th international conference on intelligent transportation systems}}. IEEE, \bibinfo{pages}{982--988}.
\newblock


\bibitem[Xu et~al\mbox{.}(2020)]%
        {xu2020adversarial}
\bibfield{author}{\bibinfo{person}{Xing Xu}, \bibinfo{person}{Jingran Zhang}, \bibinfo{person}{Yujie Li}, \bibinfo{person}{Yichuan Wang}, \bibinfo{person}{Yang Yang}, {and} \bibinfo{person}{Heng~Tao Shen}.} \bibinfo{year}{2020}\natexlab{}.
\newblock \showarticletitle{Adversarial Attack Against Urban Scene Segmentation for Autonomous Vehicles}.
\newblock \bibinfo{journal}{\emph{IEEE Transactions on Industrial Informatics}} \bibinfo{volume}{17}, \bibinfo{number}{6} (\bibinfo{year}{2020}), \bibinfo{pages}{4117--4126}.
\newblock


\bibitem[Zhang and Cai(2023)]%
        {zhang2023building}
\bibfield{author}{\bibinfo{person}{Xudong Zhang} {and} \bibinfo{person}{Yan Cai}.} \bibinfo{year}{2023}\natexlab{}.
\newblock \showarticletitle{Building critical testing scenarios for autonomous driving from real accidents}. In \bibinfo{booktitle}{\emph{Proceedings of the 32nd ACM SIGSOFT International Symposium on Software Testing and Analysis}}. \bibinfo{pages}{462--474}.
\newblock


\bibitem[Zhang et~al\mbox{.}(2023b)]%
        {zhang2023testing}
\bibfield{author}{\bibinfo{person}{Xiaodong Zhang}, \bibinfo{person}{Wei Zhao}, \bibinfo{person}{Yang Sun}, \bibinfo{person}{Jun Sun}, \bibinfo{person}{Yulong Shen}, \bibinfo{person}{Xuewen Dong}, {and} \bibinfo{person}{Zijiang Yang}.} \bibinfo{year}{2023}\natexlab{b}.
\newblock \showarticletitle{Testing automated driving systems by breaking many laws efficiently}. In \bibinfo{booktitle}{\emph{Proceedings of the 32nd ACM SIGSOFT International Symposium on Software Testing and Analysis}}. \bibinfo{pages}{942--953}.
\newblock


\bibitem[Zhang et~al\mbox{.}(2023a)]%
        {zhang2023multimodal}
\bibfield{author}{\bibinfo{person}{Zhuosheng Zhang}, \bibinfo{person}{Aston Zhang}, \bibinfo{person}{Mu Li}, \bibinfo{person}{Hai Zhao}, \bibinfo{person}{George Karypis}, {and} \bibinfo{person}{Alex Smola}.} \bibinfo{year}{2023}\natexlab{a}.
\newblock \showarticletitle{Multimodal chain-of-thought reasoning in language models}.
\newblock \bibinfo{journal}{\emph{arXiv preprint arXiv:2302.00923}} (\bibinfo{year}{2023}).
\newblock


\bibitem[Zhong et~al\mbox{.}(2022)]%
        {zhong2022neural}
\bibfield{author}{\bibinfo{person}{Ziyuan Zhong}, \bibinfo{person}{Gail Kaiser}, {and} \bibinfo{person}{Baishakhi Ray}.} \bibinfo{year}{2022}\natexlab{}.
\newblock \showarticletitle{Neural network guided evolutionary fuzzing for finding traffic violations of autonomous vehicles}.
\newblock \bibinfo{journal}{\emph{IEEE Transactions on Software Engineering}} (\bibinfo{year}{2022}).
\newblock


\bibitem[Zhou et~al\mbox{.}(2023b)]%
        {zhou2023collision}
\bibfield{author}{\bibinfo{person}{Jixiang Zhou}, \bibinfo{person}{Shuncheng Tang}, \bibinfo{person}{Yan Guo}, \bibinfo{person}{Yan-Fu Li}, {and} \bibinfo{person}{Yinxing Xue}.} \bibinfo{year}{2023}\natexlab{b}.
\newblock \showarticletitle{From Collision to Verdict: Responsibility Attribution for Autonomous Driving Systems Testing}. In \bibinfo{booktitle}{\emph{2023 IEEE 34th International Symposium on Software Reliability Engineering (ISSRE)}}. IEEE, \bibinfo{pages}{321--332}.
\newblock


\bibitem[Zhou et~al\mbox{.}(2023a)]%
        {zhou2023specification}
\bibfield{author}{\bibinfo{person}{Yuan Zhou}, \bibinfo{person}{Yang Sun}, \bibinfo{person}{Yun Tang}, \bibinfo{person}{Yuqi Chen}, \bibinfo{person}{Jun Sun}, \bibinfo{person}{Christopher~M Poskitt}, \bibinfo{person}{Yang Liu}, {and} \bibinfo{person}{Zijiang Yang}.} \bibinfo{year}{2023}\natexlab{a}.
\newblock \showarticletitle{Specification-based Autonomous Driving System Testing}.
\newblock \bibinfo{journal}{\emph{IEEE Transactions on Software Engineering}} (\bibinfo{year}{2023}).
\newblock


\bibitem[Zhou and Sun(2019)]%
        {zhou2019metamorphic}
\bibfield{author}{\bibinfo{person}{Zhi~Quan Zhou} {and} \bibinfo{person}{Liqun Sun}.} \bibinfo{year}{2019}\natexlab{}.
\newblock \showarticletitle{Metamorphic testing of driverless cars}.
\newblock \bibinfo{journal}{\emph{Commun. ACM}} \bibinfo{volume}{62}, \bibinfo{number}{3} (\bibinfo{year}{2019}), \bibinfo{pages}{61--67}.
\newblock


\bibitem[Zohdinasab et~al\mbox{.}(2021)]%
        {zohdinasab2021deephyperion}
\bibfield{author}{\bibinfo{person}{Tahereh Zohdinasab}, \bibinfo{person}{Vincenzo Riccio}, \bibinfo{person}{Alessio Gambi}, {and} \bibinfo{person}{Paolo Tonella}.} \bibinfo{year}{2021}\natexlab{}.
\newblock \showarticletitle{Deephyperion: exploring the feature space of deep learning-based systems through illumination search}. In \bibinfo{booktitle}{\emph{Proceedings of the 30th ACM SIGSOFT International Symposium on Software Testing and Analysis}}. \bibinfo{pages}{79--90}.
\newblock


\end{thebibliography}

\end{document}